\documentclass[prx,aps,epsf,twocolumn,showpacs,superscriptaddress]{revtex4-1}
\usepackage[pdftex]{graphicx}
\usepackage{dcolumn}
\usepackage{bm}
\usepackage{epsfig}
\usepackage{latexsym} 
\usepackage{amsmath}
\usepackage{amssymb}
\usepackage{color}
\usepackage{array}
\usepackage{bbm}
\usepackage{hyperref}
\usepackage{color}
\usepackage{array}
\usepackage{cancel}
\usepackage{ulem}

\usepackage{booktabs}
\newcolumntype{C}[1]{>{\centering\arraybackslash}m{#1}}
\AtBeginDocument{
\heavyrulewidth=.08em
\lightrulewidth=.05em
\cmidrulewidth=.03em
\belowrulesep=.65ex
\belowbottomsep=0pt
\aboverulesep=.4ex
\abovetopsep=0pt
\cmidrulesep=\doublerulesep
\cmidrulekern=.5em
\defaultaddspace=.5em
}
\usepackage{booktabs}
\newcolumntype{R}[1]{>{\raggedleft\arraybackslash}p{#1}}
\AtBeginDocument{
\heavyrulewidth=.08em
\lightrulewidth=.05em
\cmidrulewidth=.03em
\belowrulesep=.65ex
\belowbottomsep=0pt
\aboverulesep=.4ex
\abovetopsep=0pt
\cmidrulesep=\doublerulesep
\cmidrulekern=.5em
\defaultaddspace=.5em
}

\newcommand{\<}{\langle}
\newcommand{\e}{\varepsilon}
\newcommand{\up}{\uparrow}
\newcommand{\down}{\downarrow}
\renewcommand{\>}{\rangle}
\renewcommand{\(}{\left(}
\renewcommand{\)}{\right)}
\renewcommand{\[}{\left[}
\renewcommand{\]}{\right]}

\renewcommand{\d}{\partial}

\newcommand{\Z}{\mathbb{Z}}
\newcommand{\T}{\mathcal{T}}

\newcommand{\header}[1]{\section{#1}}

\begin{document}
\title{Dynamically enriched topological orders in driven two-dimensional systems}
\author{Andrew C Potter}
\affiliation{Department of Physics, University of Texas at Austin, Austin, TX 78712}
\author{Takahiro Morimoto}
\affiliation{Department of Physics, University of California, Berkeley, CA 94720, USA}
\begin{abstract}
Time-periodic driving of a quantum system can enable new dynamical topological phases of matter that could not exist in thermal equilibrium. We investigate two related classes of dynamical topological phenomena in 2D systems: Floquet symmetry protected topological phases (FSPTs), and Floquet enriched topological orders (FETs). By constructing solvable lattice models for a complete set of 2D bosonic FSPT phases, we show that bosonic FSPTs can be understood as topological pumps which deposit loops of 1D SPT chains onto the boundary during each driving cycle, which protects a non-trivial edge state by dynamically tuning the edge to a self-dual point poised between the 1D SPT and trivial phases of the edge. By coupling these FSPT models to dynamical gauge fields, we construct solvable models of FET orders in which anyon excitations are dynamically transmuted into topologically distinct anyon types during each driving period. These bosonic FSPT and gauged FSPT models are classified by group cohomology methods. In addition, we also construct examples of ``beyond cohomology" FET orders, which can be viewed as topological pumps of 1D topological chains formed of emergent anyonic quasi-particles.
\end{abstract}
\maketitle

\onecolumngrid
\vspace{-0.3in}
\tableofcontents
\vspace{0.3in}
\twocolumngrid

\newpage

\header{Introduction}
Periodic driving of a quantum system can be used to engineer new effective interactions that change the topological properties of a system\cite{oka2009photovoltaic,lindner2011floquet,lindner2013topological,wang2013observation}. A particularly intriguing extension of these ideas is that driving can lead to fundamentally new dynamical phases with no equilibrium counterpart\cite{kitagawa2010topological,jiang2011majorana,asboth2014chiral,gannot2015effects,thakurathi2013floquet,thakurathi2014majorana,von2016phaseI,else2016classification,potter2016topological,roy2016abelian,roy2016periodic}. Many-body localization (MBL) enables sharp distinctions between quantum-coherent dynamical phases in highly excited states\cite{huse2013localization,bauer2013area}, and allows for periodic driving of generic quantum many-body systems without heating and decoherence\cite{ponte2015many,lazarides2015fate,abanin2016theory}. Together, these developments raise the question: can we systematically understand the quantum phases and their physical properties of quantum phases in periodically driven matter? Further impetus for this study stems from recent experimental progress in producing MBL states of cold atoms\cite{schreiber2015observation,choi2016exploring,bordia2016periodically} and trapped ions\cite{smith2015many} experiments including periodically shaken\cite{bordia2016periodically} and two dimensional\cite{choi2016exploring} optical lattices.

Early investigations into this question reveal a number of intriguing new dynamical phenomena. For example, new symmetry breaking phases with dynamical order parameters may arise\cite{khemani2015phase,von2016phaseII}, including the possibility of breaking of discrete time-translation invariance without any accompanying static symmetry breaking -- a (discrete) Floquet time-crystal\cite{else2016floquet,von2016absolute,else2016pre,yao2016discrete}, which was recently observed experimentally\cite{zhang2016observation}. An additional possibility is that periodic driving enables fundamentally new topological phases\cite{rudner2013anomalous,titum2016anomalous,po2016chiral,harper2016stability}, or symmetry protected topological phases of matter\cite{kitagawa2010topological,jiang2011majorana,asboth2014chiral,gannot2015effects,thakurathi2013floquet,thakurathi2014majorana,von2016phaseI,else2016classification,potter2016topological,roy2016abelian,roy2016periodic}.

Such Floquet symmetry protected topological (FSPT) phases were originally investigated in the context of Floquet band theory relevant to weakly interacting fermions systems\cite{kitagawa2010topological,jiang2011majorana,asboth2014chiral,gannot2015effects,thakurathi2013floquet,thakurathi2014majorana,roy2016periodic}. However, most (possibly all) weakly interacting fermion SPT systems cannot be localized, and hence are unstable to heating and do not represent stable many body phases\cite{potter2016symmetry} .

Strongly interacting systems such as bosonic or spin systems, on the other hand, can avoid these problems. Obtaining non-trivial FSPT phases in such systems intrinsically requires interactions, which presents theoretical challenges. Substantial progress has been made in systematically understanding 1D FSPT phases in interacting boson systems\cite{von2016phaseI,else2016classification,potter2016topological,roy2016abelian}, where a number of intrinsically dynamical FSPT phases with no static analogs can be realized, often with rather simple interactions. Physically, these new 1D FSPT phases are characterized by a quantized amount of symmetry charge being pumped across the system and deposited during each Floquet cycle. Due to this symmetry pumping, these 1D FSPT phases exhibit protected edge states that undergo a topologically quantized spin-echo sequence, which decouples them from bulk sources of decoherence, and allows them to perfectly store quantum information. Formally, the group-structure of 1D FSPT phases are classified by projective representations of the symmetry group extended by a discrete time-translation symmetry due to the Floquet structure\cite{else2016classification,potter2016topological}. For bosonic systems, such projective representations are encoded in the second group cohomology\cite{chen2012symmetry,chen2013symmetry,else2016classification,potter2016topological}.

In this paper, we focus on understanding FSPT phases of bosonic systems (e.g. spin systems) in two and higher dimensions. A natural extension of of the 1D FSPT classification to arbitrary $d$ dimensional systems, is to consider the the $(d+1)^\text{th}$ group cohomology of the on-site symmetries plus a discrete time-translation symmetry. This classification was proposed in \cite{else2016classification,potter2016topological}, and physically interpreted as being characterized by a quantized pumping of $d-1$ dimensional SPT phases across the system for each Floquet cycle\cite{else2016classification}. Further, exactly solvable models of the new dynamical phases were sketched in \cite{else2016classification}. 

While these results derive the abstract group structure of higher dimensional bosonic FSPTs, and offer a proof of principle that they can arise in local spin models, the physical properties of these new phases remain to be elucidated. In particular, as with any SPT phase, we expect the FSPT phases to exhibit unconventional edge states with anomalous implementations of symmetry. In equilibrium systems, the 1D edges of 2D bulk SPTs must either spontaneously break the protecting symmetry, or exhibit symmetry protected gapless modes with anomalous implementations of symmetry. What are the analogous sets of edge phases for non-equilibrium periodically driven FSPTs? For example, one might expect that under periodic driving, the gapless surface state of a bulk SPT will absorb energy and become ergodic, heating to infinite temperature. Is this necessarily the fate of the symmetry-preserving edges of Floquet SPTs? Can the anomalous properties of the edge still be observed at infinite temperature?

To address such questions, we construct a simple, solvable model of a 2D Floquet SPT with $\Z_2\times\Z_2$ symmetry -- which can be viewed as a phase in which a 1D SPT with the same symmetry is pumped across the system onto the edge during each Floquet period. We explicitly analyze the edge of this theory, and elucidate how the bulk pumping action dynamically enlarges the symmetry group of the edge to a non-Abelian symmetry, which prevents the edge from localizing while preserving symmetry. For the set of edge Hamiltonians considered, we find that the edge of this 2D FSPT phase can either be symmetry preserving and thermal or symmetry-breaking and localized. We then generalize this construction to other bosonic 2D FSPT classes. We describe edge and bulk diagnostics of the dynamical SPT orders, including signatures in the micro-motion of the entanglement spectrum. Interestingly, we show that the dynamical anomalous properties of the edge survive even when the edge is thermal and heats up to infinite temperature by absorbing energy from the drive.

Next we investigate new dynamical phases of periodically driven systems with intrinsic topological order and fractional anyonic excitations. A la \cite{levin2012braiding}, we start by promoting the global protecting symmetry of a bosonic FSPT models to a local gauge symmetry, to obtain new Floquet enriched topological phases (FETs). Like the corresponding 2D bosonic FSPT models, this class of FETs are classified by group cohomology\cite{dijkgraaf1990topological} with an extra dynamical symmetry corresponding to time-translation symmetry. However, we show that such cohomology states do not exhaust the possible set of FETs, and construct several examples of ``beyond cohomology" FET phases. These examples can be considered as anyonic generalizations of the 2D FSPTs, in which 1D topological phases of anyonic quasiparticles are pumped onto the edge of the system during each driving period. An alternative way to characterize these phases is that the bulk anyon types are dynamically permuted during each driving period. 
We find that, in the MBL regime, the dynamical permutation of anyons forces the FET phase to be accompanied by spontaneous time-crystal order, but that the FET order remains sharply defined despite the spontaneously enlarged periodicity in the time-crystal phase.

In addition to these FSPT phases, there are possible 2D chiral Floquet phases\cite{rudner2013anomalous,titum2016anomalous,po2016chiral} characterized by quantized chiral pumping of quantum information at the edge\cite{po2016chiral}, and which do not rely on symmetry protection\cite{po2016chiral}. In this work, we will restrict our attention to non-chiral Floquet phases, and leave the challenge of incorporating chiral edge pumping into the classification of FSPT and FET phases for future work. We also do not consider the more challenging problem of describing Floquet enrichment of non-Abelian topological orders, which can be expected to lead to a rich variety of new FET and chiral Floquet phases.

\section{Review: FSPT classification in one and higher dimensions}
The 1D bosonic FSPTs with symmetry group $G$ are classified by projective representations of the an enlarged symmetry group, $\tilde{G}$, that includes both the symmetries of the Floquet Hamiltonian, $G$, together with an extra discrete time translation symmetry which gives an extra factor of $\Z$: $\tilde{G}=\Z\times G$ (or $\Z\rtimes G$), for $G$ unitary (antiunitary) respectively. The natural generalization of this classification to higher dimensions is that the group structure of $d$-dimensional bosonic FSPT phases is given by the $d+1$ group-cohomology:
\begin{align}
\mathcal{H}^{d+1}(G\times \Z,U(1))=
\underset{\text{static SPTs}}{\underbrace{\mathcal{H}^{d+1}(G,U(1))}}
\times
\underset{\text{dynamical FSPTs}}{\underbrace{\mathcal{H}^d(G\times \Z,U(1))}}
\end{align}
In the second line, we have applied the Kunneth formula\cite{chen2013symmetry,else2016classification,potter2016topological}, which reveals that the FSPT phases decompose into those with static SPT order that could also occur in zero temperature equilibrium systems, and extra intrinsically dynamical FSPT orders which can only occur in driven systems. 

In 1D, the classification of the extra dynamical phases corresponds to $\mathcal{H}^1(G,U(1))$ -- i.e. to representations of the symmetry $G$ \cite{von2016phaseI,else2016classification,potter2016topological}. The physical picture of these extra dynamical FSPT phases, is that the Floquet drive induces a quantized pumping of symmetry charge across the system during each Floquet period. This pumping protects localized edge degrees of freedom (``spins") that can freely flip to absorb the pumped charge. Intuitively, the flipping of the edge spins produces a topologically robust spin-echo that dynamically decouples them from bulk degrees of freedom. Different SPT phases correspond to different quantized amount of charge pumped per period, and the distinct possible charges correspond precisely to the distinct representations of $G$.

What is the analogous physical picture for higher dimensional Floquet SPTs? For $d>1$, the extra factor in the cohomology classification associated with dynamical FSPT phases correspond to static SPTs in one lower dimension. By analogy to the symmetry charge-pumping in 1D, Else and Nayak\cite{else2016classification} proposed that the new example phases could be thought of as dynamically pumping a lower-dimensional SPT onto the edge during each Floquet period. However, from this construction, it is not immediately clear how this pumping dynamically protects nontrivial SPT edge states, or how to characterize the physical properties of various possible edge phases.

\section{2D bosonic FSPT with $\Z_2\times\Z_2$ symmetry}
To better understand the physical properties of the proposed higher dimensional FSPT phases, we construct a concrete time-dependent Hamiltonian for a 2D FSPT, in which a 1D SPT (a discrete analog of the Haldane spin-chain\cite{haldane1983nonlinear,AKLT}) protected by $G=\Z_2\times\Z_2$ symmetry is pumped onto the boundary during each period. From this construction, we can directly analyze various possible boundary phases.

\subsection{Solvable spin model}
A crucial building block for the ``pumping" in the two dimensional model is a unitary operator that acts on a closed 1D spin chain, and interchanges the 1D SPT phase with $\Z_2\times \Z_2$ symmetry, with a trivial paramagnet of the same symmetry. Namely, consider a 1D spin chain with 4-states per site, which we will decompose into two spin-1/2 operators living on an A and B sub-lattice, where the separate $\Z_2$ factors of $G$ are generated by flipping the spins on either the A or B sublattice: $g_{A,B} = \prod_{i\in A,B}\sigma_i^x$.  With this symmetry, there is a single non-trivial static SPT phase, exemplified by the ground state of the Hamiltonian:
\begin{align}
H_\text{1D-SPT} = \sum_i \lambda_i \sigma^z_{i-1}\sigma^x_i\sigma^z_{i+1} 
\end{align}
For periodic boundary conditions, this SPT Hamiltonian is mapped into the trivial paramagnet $H_\text{1D-PM} = \sum_i \lambda_i\sigma_i^x$ by the unitary operator:
\begin{align}
U_\text{SPT} = e^{i\frac{\pi}{4}\sum_i (-1)^i\sigma^z_i\sigma^z_{i+1}}
\end{align}
which one can readily verify interchanges $\sigma_{i-1}^z\sigma_i^x\sigma_{i+1}^z\leftrightarrow \sigma_{i}^x$ (the alternating $(-1)^i$ factor is chosen to avoid an unwanted overall phase). The operator $U_\text{SPT}$ preserves the symmetry generators, however the related family of operators $e^{i\pi\lambda/4\sum_i\sigma_i^z\sigma_{i+1}^z}$, which interpolates between the identity and $U_\text{SPT}$ for $\lambda \in [0,1]$ does not preserve the symmetry except at the endpoints $\lambda=0,1$. Alternatively, we may define a symmetry-preserving, but non-local Hermitian operator $H_\text{pump}=i\log U_\text{SPT}$, which, when exponentiated as $U_\lambda = e^{-i\lambda H_\text{pump}}$, continuously interpolates between the identity and $U_\text{SPT}$ for $\lambda\in [0,1]$. The price of making $U_\lambda$ symmetry preserving for all $\lambda$, is that $H_\text{pump}$ is non-local, including products of arbitrarily long strings of spin-operators. 

While non-local on an infinite chain, for a short 4-site loop, with sites $1,2,3,4$, and periodic boundaries, $H_\text{pump}$ obtains a particularly simple form:
\begin{align}
U^{(1234)}_\text{SPT}&=\exp\[i\frac{\pi}{4}\sum_{i=1}^4 (-1)^i\sigma^z_{i}\sigma^z_{i+1\text{mod}4}\] 
\nonumber\\
&= \frac{1}{2}\(1+\sigma^z_1\sigma^z_2\sigma^z_3\sigma^z_4-\sigma^z_1\sigma^z_3-\sigma^z_2\sigma^z_4\)
\nonumber\\
&= \exp\[i\pi\(\frac{1-\sigma^z_1\sigma^z_3}{2}\) \(\frac{1-\sigma^z_2\sigma^z_4}{2}\)\] 
\end{align}
from which we read off: $H^{(1234)}_\text{pump} = \frac{\pi}{4}\(1- \sigma^z_1\sigma^z_3\)\(1- \sigma^z_2\sigma^z_4\)$. This term can be physically interpreted as a time evolution governed by a short-range interaction between domain walls in the $\Z_2^A$ and $\Z_2^B$ symmetry breaking orders.

\begin{figure}[t!]
\includegraphics[width=\columnwidth]{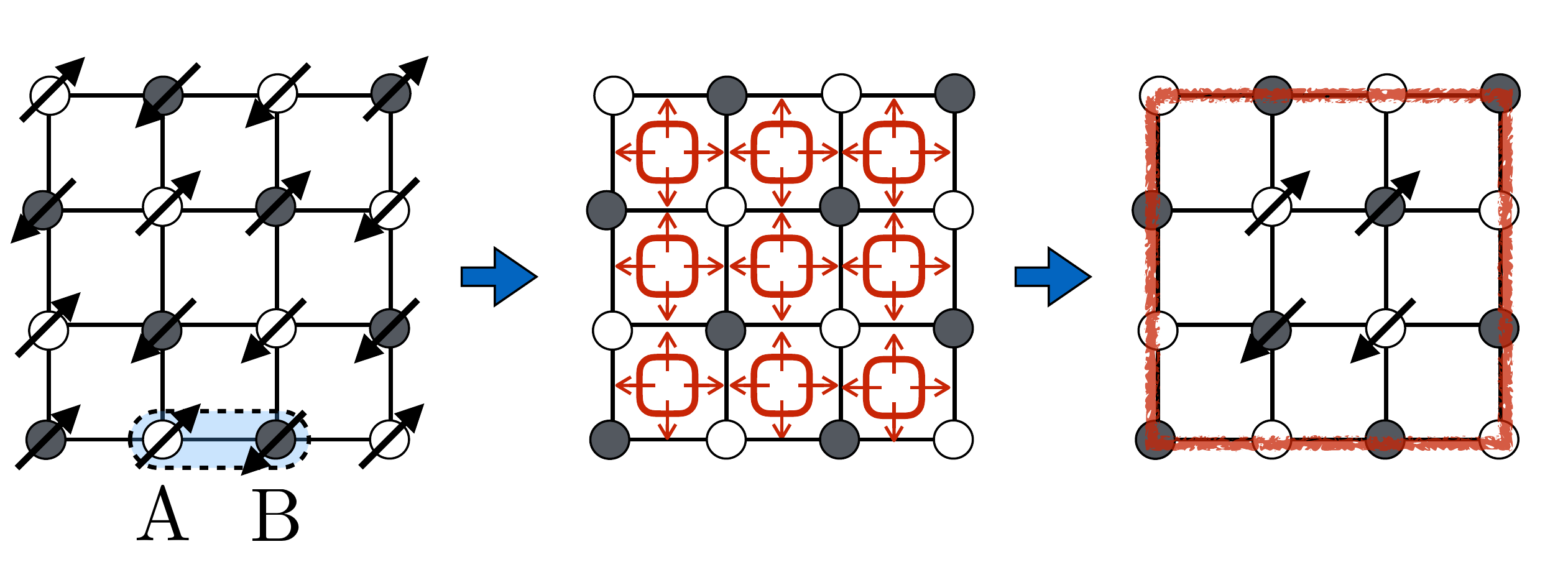}
\caption{ {\bf Model - } The spin model consists of Ising spins on a square lattice, with a two site (A and B) unit cell (dashed shaded oval in left panel). The sequence schematically depicts how locally pumping a short loops of 1D SPTs onto each plaquette results in toggling the edge between trivial unentangled state and the 1D SPT phase. This dynamically tunes the edge to a self-dual point as described in the text. }
\label{fig:pumping}
\end{figure}

With these ingredients in hand, we are ready to construct a time-dependent model Hamiltonian to realize the 2D FSPT phase in which this 1D static SPT is pumped onto the boundary during each cycle. Consider a 2D square lattice with 4-state spins on each site, which we can describe as spins-1/2 living on the A and B sub-lattices (Fig.~\ref{fig:pumping}), again with $\Z_2^A\times\Z_2^B$ symmetry generators $g_{A/B}=\prod_{i\in A/B}\sigma_i^x$. Let us subject this system to a piecewise constant (``stroboscopic") time-dependent Hamiltonian:
\begin{align}
H(t) = 
\begin{cases}
2H_1 = 2\sum_P H^{(P)}_{\text{pump}} & 0\leq t< 1/2
\vspace{4pt} \\ 
2H_2 = 2\sum_i h_i\sigma_i^x & 1/2\leq t<1
\end{cases}
\label{eq:HZ2xZ2}
\end{align}
where we have normalized the time by the Floquet period, $T\equiv 1$, and arranged the factors of $2$ such that $U(T) = e^{-iH_2}e^{-iH_1}$. The first term includes a sum over all 4-spin plaquettes, $P$, on the square lattice. Each of these plaquette terms has the effect of driving the boundary spins of the plaquette between the locally SPT and trivial phases -- i.e. of pumping a short loop of 1D SPT onto the boundary of each plaquette. 

In the bulk, each link of the lattice borders two plaquettes, and hence after one period has two SPT edges pumped onto it. Since two copies of the 1D SPT in question are topologically trivial, the effect of $H_1$ after evolution by half a period is trivial in the bulk. However, at the system's boundary, each edge link neighbors only one plaquette, and is hence toggled between the topological and trivial phases over the course of each Floquet period -- producing the desired pumping of the 1D SPT onto the boundary. The second stage of the Hamiltonian, $H_2$, simply represents time-evolution by a trivial paramagnet, for which sufficiently random transverse fields, $h_i$, ensure many-body localization. I.e. in total, the Floquet evolution operator acting on a system occupying spatial region $\Omega$ for a full period reads:
\begin{align}
U(T) = e^{-i\sum_{i\in \Omega} h_i\sigma_i^x} U_{\text{SPT},\d \Omega} 
\end{align}
where $U_{\text{SPT},\d\Omega}$ pumps a 1D SPT phase onto the boundary $\d\Omega$.

\subsection{Edge State Properties} 
We now turn to an analysis of the edge properties of the above model. For static, equilibrium 2D SPT states, the edge is either gapless or symmetry breaking. Similarly, in non-driven bulk MBL systems, the edge of a 2D SPT cannot be both MBL and symmetry preserving. What is the analogous situation for 2D Floquet SPT edges? One expects the dynamics to play a crucial role in protecting the edge. By analyzing the edge of the 2D FSPT with $\Z_2\times\Z_2$ symmetry, we will see that there is an emergent dynamical $\Z_2$ symmetry associated with time-translation that does not commute with the other symmetry generators, and effectively promotes the action of symmetry at the edge to a non-Abelian dihedral symmetry. Recent work\cite{potter2016symmetry,vasseur2015quantum,vasseur2015particle} shows that systems with non-Abelian symmetries cannot be trivially localized without spontaneously breaking the symmetry. In brief, in an MBL system where all excitations are local, symmetry also acts locally. Since non-Abelian symmetries protect degenerate multiplets, in the MBL context, local excitations would form these exactly degenerate multiplets. This situation is inherently unstable to generic perturbations, which induce exactly resonant mixing of the local multiplets, resulting in either delocalization or the spontaneous breaking of the symmetry to lift the degeneracy. In the present context, we connect this non-Abelian obstruction to symmetry-preserving localization to the symmetry and topological protection of the 2D FSPT edge.

\subsubsection{Anomalous dynamical edge symmetry}
A technical difficulty in analyzing the Floquet evolution at the edge, is that the operator $U(T)$ does not take the form of time evolution with a local, symmetry preserving Hamiltonian: $U(T)\neq e^{-iH_\text{loc}}$, due to the $U_\text{SPT}$ term. To avoid this, we may follow the approach of \cite{yao2016discrete}, and instead look at the Floquet evolution operator for two periods:
\begin{align}
U(2T)_\text{edge} &= \(e^{-iH_{2}}U_\text{SPT}\)^2 = e^{-iH_2}e^{-iU_\text{SPT}H_2U_\text{SPT}^\dagger}
\end{align}
which we will see shortly, does take the form of a local symmetry-preserving Hamiltonian evolution: 
\begin{align}
U(2T)_\text{edge} = e^{-2iH_\text{eff}}
\end{align}
where $H_\text{eff}$ is local (i.e. its terms are exponentially well localized) and preserves symmetry. Note that, generally, if we consider a phase in which a 1D SPT with a $\Z_n$ topological index is being pumped, then to obtain a local effective Hamiltonian description of the edge dynamics, we should consider $U(nT)_\text{edge}$.

In considering $U(2T)$ we are implicitly considering stroboscopic measurements at integer multiples of twice the driving period. However, the $U(2T)$ retains an imprint of the fact that the system is subjected to a fundamental $T$-periodic drive. Namely, there is an emergent dynamical $\Z_2$ symmetry associated with discrete time translation generated by\cite{yao2016discrete}:
\begin{align}
g_D = U(T)e^{iH_\text{eff}}
\end{align}
By construction, this dynamical symmetry generator: i) commutes with $U(T)$ (i.e. is a symmetry of the dynamics), ii) obeys a $\Z_2$ multiplication rule $g_D^2=1$, and iii) is the product of (quasi)-local unitary operators (though not generated by exponentiating a symmetry preserving Hamiltonian). We dub this symmetry ``emergent" as the explicit form of its generator depends on the terms in the Floquet Hamiltonian. Roughly speaking, we can view this symmetry as isolating the ``interesting" part of the dynamics of $U(T)$ that is effected by the SPT pumping from the bulk, from the trivial (non-anomalous) dynamical phases acquired from surface excitations that are unaffected by the SPT pumping. To see this, note that any terms in $H_2$ that commute with $U_\text{SPT}$ are cancelled from $g_D$ by the $e^{iH_\text{eff}}$ factor.

The dynamical symmetry schematically takes the form $g_D = \(\prod_i e^{i\frac{\pi}{4} \(-1\)^i \sigma^z_i\sigma^z_{i+1}}\)\(\dots\)$, where $(\dots)$ represents terms that commute with the symmetry, and we have explicitly written the effect of $U_\text{SPT}$ at the edge in terms of the pumping Hamiltonian $H_\text{pump}$. We see that while the global action of time-evolution commutes with that of the symmetry generators at the edge, the local action of symmetry does not commute with $g_D$: $g_D\sigma_i^xg_D^\dagger = \sigma_{i-1}^z\sigma^x_i\sigma_{i+1}^z$. This failure to locally commute will place strong constraints on MBL systems, in which excitations are local and transform under the local action of symmetry.

To investigate further, we consider a concrete form of the edge dynamics, with a random transverse field $H_2 = \sum_i h_i\sigma_i^x$. Moreover, to facilitate closed form expressions we will work in the high-frequency limit $h_i\ll 2\pi$. In this limit, one finds that:
\begin{align}
H_\text{eff} &= \frac{i}{2}\log U(2T)_\text{edge}
\nonumber\\ 
&\approx \sum_{i\in \text{edge}} \frac{H_2+\tilde{H}_2}{2}+\frac{-i}{4}\[H_2,\tilde{H}_2\]+\mathcal{O}(H_2^3)
\end{align}
where $\tilde{H_2} = U_\text{SPT}H_2U_\text{SPT}$. One can also work out the explicit form of the dynamical symmetry generator to the same order in the high-frequency expansion: 
\begin{align}
g_D\approx e^{iH_2/2}U_\text{SPT}e^{-iH_2/2}+\mathcal{O}(H_2^3)
\end{align}
Note that the dynamical symmetry is related to $U_\text{SPT}$ by a finite-depth (Hamiltonian dependent) transformation. Generically, we expect that this feature will hold exactly for MBL edge Hamiltonians.

We can substantially simplify all expressions by going into a rotated basis: $|\Psi\>\rightarrow e^{-iH_2/2}|\Psi\>$, in which the dynamical symmetry and effective edge Hamiltonian become simply:
\begin{align}
g_D\rightarrow g_D' = U_\text{SPT}+\mathcal{O}(H_2^3)
\end{align}
Note also, in this rotated basis the commutator terms drop out of the effective edge Hamiltonian, which simply becomes: 
\begin{align}
H_\text{eff}\rightarrow H'_\text{eff}=\sum_i\frac{h_i}{2}\(\sigma^x_i+\sigma^z_{i-1}\sigma^x_i\sigma^z_{i+1}\) +\mathcal{O}(h)^3
\end{align}

This effective edge Hamiltonian is automatically tuned, by the 1D SPT pumping action of the bulk, to the self dual point of the duality transformation: 
\begin{align}
\tilde\sigma^x_i = \sigma^z_{i-1}\sigma^x_i\sigma^z_{i+1}
\nonumber\\
\tilde\sigma^z_{i-1}\tilde\sigma^x_{i}\tilde\sigma^z_{i+1} = \sigma^x_i
\end{align}
which interchanges the 1D SPT ordered and trivial phases of the edge.

Note that the dynamical symmetry is equivalent to the action of $U_\text{SPT}$, up to a $\Z_2\times\Z_2$ symmetry preserving finite depth unitary transformation (in this case spatially random $x$-rotations). In the ground state of $H_\text{eff}$ this self-duality would ensure that the edge is tuned to a quantum phase transition separating these two phases. Since energy is not conserved in a Floquet system, we are instead interested in generic excited states of $H_\text{eff}$, whose behavior turns out to be substantially different.

To proceed, it is useful to rewrite $H_\text{eff}$ in terms of dual spins $\vec{\tau}$ residing on the bonds of the original lattice, defined as:
\begin{align}
\sigma^x_i\rightarrow \tau_{i-1/2}^x\tau_{i+1/2}^x 
\nonumber\\
\sigma^z_{i-1}\sigma^x_i\sigma^z_{i+1} \rightarrow \tau^y_{i-1/2}\tau^y_{i+1/2}
\end{align}
In terms of these new variables, the effective edge Hamiltonian takes the form of a random bond XX chain:
\begin{align}
H_\text{eff}' = \sum_i h_i \(\tau^x_{i-1/2}\tau^x_{i+1/2}+\tau^y_{i-1/2}\tau^y_{i+1/2}\)
\end{align}
Here, we notice an enlarged $U(1)$ symmetry generated by rotations around $\tau^z$: $R(\alpha) = e^{-i\frac{\alpha}{2}\sum_i\tau^z_{i+1/2}}$. This symmetry was not manifest in the original model, however, we can directly trace a sub-group of this $U(1)$ symmetry to the dynamical symmetry $g_D$. Namely, inverting the above duality transformation, one can write: $\tau^z_{i-1/2} = (-1)^i\sigma^z_{i-1}\sigma^z_{i}$, from which we immediately see that $\frac{\pi}{2}$ rotations implement the dynamical pumping of the 1D SPT on the edge: $R(\pi/2) = g_D$. Hence we expect that with arbitrary interactions only this $\Z_4$ subgroup will survive. However, note that at the level of two-spin, nearest neighbor interactions there are no additional terms we can include that break the $U(1)$ down to $\Z_4$. The simplest such terms come from either further-nearest neighbor interactions of the $\sigma$'s. Hence for the moment we will stick with the model with the enlarged symmetry. 

Note further, that the original static $\Z_2\times\Z_2$ symmetry generators both act as $\prod_{i}\sigma_{2i}^x = \prod_i\tau_{i+1/2}^x = \prod_i\sigma_{2i+1}^x$\footnote{Note that this coincidence of $\Z_2^{A/B}$ comes from a global ambiguity between the state of spins, $\sigma$, and a given configuration of ``domain wall" variables $\tau$ in a finite system, however the dynamics of the resulting phase occur locally, and are insensitive to this discrepancy}, i.e. act as a particle-hole symmetry, $\mathcal{C}$, for the $U(1)$ variables, so that the total symmetry group of $H_\text{eff}$ is then $U(1)\rtimes \mathcal{C}$. With generic interactions this will be broken down to a dihedral symmetry $\Z_4\rtimes \Z_2$. 

Hence we see that the 1D SPT pumping has the effect of dynamically promoting the Abelian bulk symmetry group to an effective non-Abelian symmetry at the edge. This non-Abelian symmetry action immediately rules out many-body localization at the edge, unless it is spontaneously broken\cite{potter2016symmetry}. Notably, the effective non-Abelian action of symmetry at the edge also rules out the possibility of a symmetric critical edge, as would happen in the ground-state. For example, there are quantum critical analog of MBL, with an extensive set of conserved quantities, but which are only algebraically well localized, as happens for instance in the strongly random critical Ising model\cite{vosk2014dynamical,pekker2014hilbert,vasseur2015quantum}, or in the edge of 2D ground-state SPTs. Even such long-range entangled but non-thermal critical points are forbidden from occurring (without accompanying symmetry breaking) in the non-equilibrium Floquet setting\cite{potter2016symmetry}. While we have demonstrated this effective non-Abelianization of the dynamical edge symmetry for a specific model, this feature is general to all 2D bosonic FSPTs systems characterized by a dynamical pumping of a lower dimensional SPT chain\cite{chen2013symmetry} -- in these phases, the dynamical symmetry plays the role of an effective $\mathbb{Z}_n$ symmetry whose domain walls carry projective (effectively non-Abelian) representations of the static symmetry group.

To analyze the allowed possibilities for the edge, let us first note that the random bond XX spin-chain is actually integrable (equivalent to free fermions), and has the pathological feature that its many-body excited states are infinitely degenerate. In the fermion language, this degeneracy arises from the equal energy obtained by filling both an orbital and its particle hole conjugate orbital, or leaving both empty. In the spin model at strong disorder, this degeneracy occurs when neighboring spins are locked into a ferromagnetic arrangement $|\up_i\up_{i+1}\>$ or $|\down_i\down_{i+1}\>$, which have equal energy. To cure this problem, we must include additional interactions, the simplest being of the form $\tau^z\tau^z$, corresponding to next-nearest neighbor interactions, $\sigma_{i-1}^z\sigma_{i+1}^z$, in the original language. The fate of excited state dynamics in such random bond XXZ model was recently studied in \cite{vasseur2015particle}, where it was found that, at strong disorder, the interactions immediately freeze the degenerate particle-hole degrees of freedom into a random product state, or equivalently a spontaneous formation of spin-glass order in $\tau^z$, which breaks the particle-hole symmetry $\mathcal{C}$ (in the $\tau$ language), or equivalently the $\Z_2\times\Z_2$ symmetry (in the $\sigma$ language).

On the other hand, the dynamical $U(1)$ symmetry remains unbroken, and hence respects the time-translation symmetry. Indeed it would not be possible for a non-equilibrium 1D system to spontaneously break this a continuous symmetry, as this would require a Goldstone mode, which would ensure thermalization, which is inconsistent since thermal systems cannot spontaneously break continuous symmetries.

\subsubsection{Other possible edge phases}
A second possibility, which is not naturally realized by having simply random transverse fields in $H_2$ at the edge, but which is in principle realizable by more complicated interactions, is to have a spontaneous breaking of time-translation symmetry alone, without an accompanying breaking of $\Z_2\times\Z_2$ symmetry\cite{else2016floquet}. In this case, with the time-translation symmetry breaking, we may spontaneously tip the balance from the critical point between SPT and trivial edge states, to form a localized edge SPT or localized edge PM phase. For example, if we either spontaneously generate or explicitly apply an effective period-$2T$ edge field $\delta h(t)\sigma^x$ which has $\delta h(t)>0$ for $0<t<T$ and $\delta h(t)<0$ for $T<t<2T$, then with statistically homogeneous edge transverse fields in $H_2$, we obtain $H_\text{eff} \approx \sum_i (h_i+\delta h)\sigma^z_{i-1}\sigma^x_{i}\sigma^z_{i+1}+(h_i-\delta h)\sigma^x_{i}$, which pushes the edge into the localized SPT phase for $\delta h>0$ or the trivial PM phase for $\delta h<0$, both of which are many-body localized and boundary-law entangled despite preserving the on-site $\Z_2\times\Z_2$ symmetry. Therefore, we see that the periodicity of the Floquet drive plays the role of an additional discrete time-translation symmetry in protecting the edge states of these dynamical FSPT phases.

Alternatively, while for strongly disordered edge couplings a symmetry-broken MBL edge phases, at weak disorder, the edge may simply thermalize. Generically, an isolated, driven thermal system will absorb energy and heat up to infinite temperature, destroying underlying quantum coherence. 
Since thermal phases in 1D necessarily preserve the $\Z_2\times\Z_2$ symmetry, we expect that the edge anomaly is still accessible even for clean, thermalizing edges. We will verify this expectation below in Sec.~\ref{sec:anomaly}, where we discuss physical characterizations of this edge anomaly for symmetry preserving edges.

Finally, we note that it is possible to have quantum critical states that are neither thermal nor MBL, exhibiting logarithmic scaling of entanglement, algebraically decaying correlation functions, and lacking a set of independent conserved quantities\cite{vasseur2015quantum,kang2016universal}. While we are unable to find such a critical, symmetry preserving edge state for a concrete edge Hamiltonian, we do not know a fundamental reason that precludes its occurrence in more complicated Hamiltonians.

\subsection{Other 2D bosonic FSPTs}
We can generalize the above construction to other symmetry classes. 
A very simple extension, is that the same $\Z_2\times\Z_2$ Hamiltonian described above, also moonlights as a model Hamiltonian for a time-reversal symmetry protected FSPT, with time-reversal symmetry acting as $\T = \prod_i\sigma_i^x K$, where $K$ denotes complex conjugation. 

For unitary symmetries, the most general symmetry group that is compatible with many-body localization\cite{potter2016topological,potter2016symmetry} are finite Abelian groups which can be written as product of $G=\prod_I \Z_{n_I}$. The 1D static SPTs for such a finite-Abelian $G$ are given by: $\prod_{I,J} Z_{\text{gcd}(n_I,n_J)}$, where $\text{gcd}$ indicates the greatest common divisor. These phases can be thought of as having quantum proliferated domain walls of the $\Z_{n_I}$ order are bound to $Z_{n_J}$ symmetry charge. Within the proposed cohomology classification of 2D FSPTs, the intrinsically dynamical 2D FSPT phases correspond to phases in which these 1D static SPTs are pumped onto the boundary. Due to the direct product structure of the 1D SPT classification, it is sufficient to consider just a single pair of symmetries: $G=\Z_n\times\Z_n$. We can readily generalize the above construction for $\Z_2\times\Z_2$ variables, in terms of $\Z_n$ valued spins (see Appendix~A). Here, there are $n$ distinct phases corresponding to pumping an integer number $j=0,1,2,\dots,n-1$. Just as in the $\Z_2\times\Z_2$ case described above, the edge of these phases can either be symmetry broken and localized or symmetry preserving and thermal.

\section{Edge and bulk diagnostics of FSPT order \label{sec:anomaly}}
In this section, we consider physical edge- and bulk- diagnostics that can be used to measure (e.g. numerically) the bulk FSPT order in a given model. We begin by discussing edge-based characterizations. As previously remarked above, based on intuition developed for static SPTs, one expects the anomalous properties of the FSPT bulk to imprint itself on the full spectrum of edge excitations, such that the edge manifests the bulk FSPT order so long as the edge preserves the underlying symmetry. In particular, we expect the FSPT order to be, in principle, measurable, even when the edge is thermalizing and heats to infinite effective temperature due to the periodic drive. 

To see this, let us consider a cylinder of the dynamical 2D $\Z_2\times\Z_2$ FSPT. Imagine inserting a $\pi$-flux defect of the $\Z_2^B$ symmetry into the access of the cylinder. Due to the dynamical 1D SPT pumping, and the fact that the 1D SPT develops $\Z_2^A$ charge upon encircling a $\Z_2^B$-flux, this $B$-symmetry defect will lead to a quantized dynamical $A$-charge-pumping on each end of the cylinder. Namely, consider measuring the change, $\Delta Q_A$, in $\Z_2^A$ charge on one edge of the cylinder in the course of one Floquet period, we will find that $\Delta Q_A$ with a $B$-symmetry flux differs by $(-1)$ compared to the case without the $B$-symmetry flux threading the cylinder. This symmetry flux-induced charge pumping provides a robust edge-signature of the 2D FSPT phase, that is present even when the edge is thermal and heats to infinite temperature. 

We note a technical caveat to keep in mind in the above thought-experiment: in the limit of infinite edge length, a thermalizing edge has a continuous many-body spectrum and can act as a bath to thermalize bulk degrees of freedom. However, for a well-localized bulk, the time-scale for this thermalization scales like $\tau_\text{therm}\sim e^{L_\perp}$, where $L_\perp$ is the length of the system. Hence, what we have in mind in this construction is a limiting procedure in which we take the $L_\perp$ to infinity parametrically parametrically faster than the length of the edge $L_\parallel$ (e.g. as $L_\perp\sim L_\parallel^2$). In this way, the time scale for the bulk thermalization diverges much faster than the edge-thermalization time-scale, so that there is a well-defined infinite system long time limit in which the edge is in a thermal steady state, but the bulk is still localized (see also \cite{chandran2016many}).

A closely related signature of the bulk FSPT order is also present in the ``micro-motion" of the entanglement spectrum of a one-dimensional cut through the system\cite{potter2016topological}. Namely, while the entanglement spectrum is $T$-periodic, the Schmidt states corresponding to a given entanglement energy change their 1D SPT invariant under each Floquet cycle, by interchanging with another state with a different 1D SPT order at some time within the Floquet period.\cite{potter2016topological}

\section{Gauging a bosonic FSPT}
For equilibrium SPTs protected by an on-site unitary symmetry group $G$, promoting the global symmetry group $G$ to a local gauge symmetry by coupling the local gauge charge to a discrete lattice gauge field (or ``gauging the symmetry") produces a symmetry enriched topological phase (SET) -- i.e. one that is distinct from gauging a trivial paramagnet. Gauging distinct SPT phases produce distinct SET phases with the same topological order but different symmetry action. In particular, there is a direct connection between the group cohomology formulation used by Dijkgraaf-Witten to classify gauge theories\cite{dijkgraaf1990topological}, and that used to classify bosonic SPT phases.

This naturally leads to the question: what dynamical topologically ordered phase results from gauging the symmetry of an intrinsically dynamical FSPT? We will generally find that, in accordance with the intuition from equilibrium phases, gauging a dynamical FSPT produces an intrinsically non-equilibrium dynamical topological order with no equilibrium counterpart. We dub such phases $\underline{\text{F}}$loquet $\underline{\text{E}}$nriched $\underline{\text{T}}$opological orders, or FETs. Here we have dropped ``symmetry" from the label for this phenomena, as we will consider the effect of periodically driving pure topological orders without any symmetry enrichment (though in a sense, the periodicity of the Floquet drive may be considered a symmetry). We note that symmetry and periodic driving could both interplay to bring about  ``Floquet and symmetry enriched topological orders" or FSETs, however, we will not consider such phases here, and instead restrict to purely dynamical enrichment of topological orders. 

Interestingly, while gauging bosonic FSPTs provides one route to constructing new FET phases, we will find that such examples do not exhaust the possible FET phases, and will construct FET phases that lie beyond the cohomology classification. 

In all cases, the FET phases we will encounter are characterized by a dynamical permutation of anyon types, in which an anyon excitation is morphed into a different type of anyon during the Floquet drive. In the fine-tuned limit of zero correlation length (perfect localization), this anyon permutation dynamically superposes different types of anyon excitations, morphing them into new objects with effective quantum dimension larger than one (though still integer), and effectively non-Abelian fusion properties. This enhanced quantum dimension enforces an extensive degeneracy of excited eigenstates, which will be lifted in a non-trivial way upon going away from the fine-tuned zero correlation length limit. As a result, we will see that the notion of FET orders in the MBL regime generically are accompanied by spontaneous time-translation symmetry breaking, and do not have topologically protected non-MBL edge states. 


For this reason, we will initially describe the properties of FET orders in the zero-correlation length (infinite disorder) limit for a variety of systems, and then move away from this idealized limit to analyze the metastable decay and eventual fate of the FET order.

\subsection{Zero-correlation length Lattice model} 
\begin{figure}[t!]
\includegraphics[width=0.7\columnwidth]{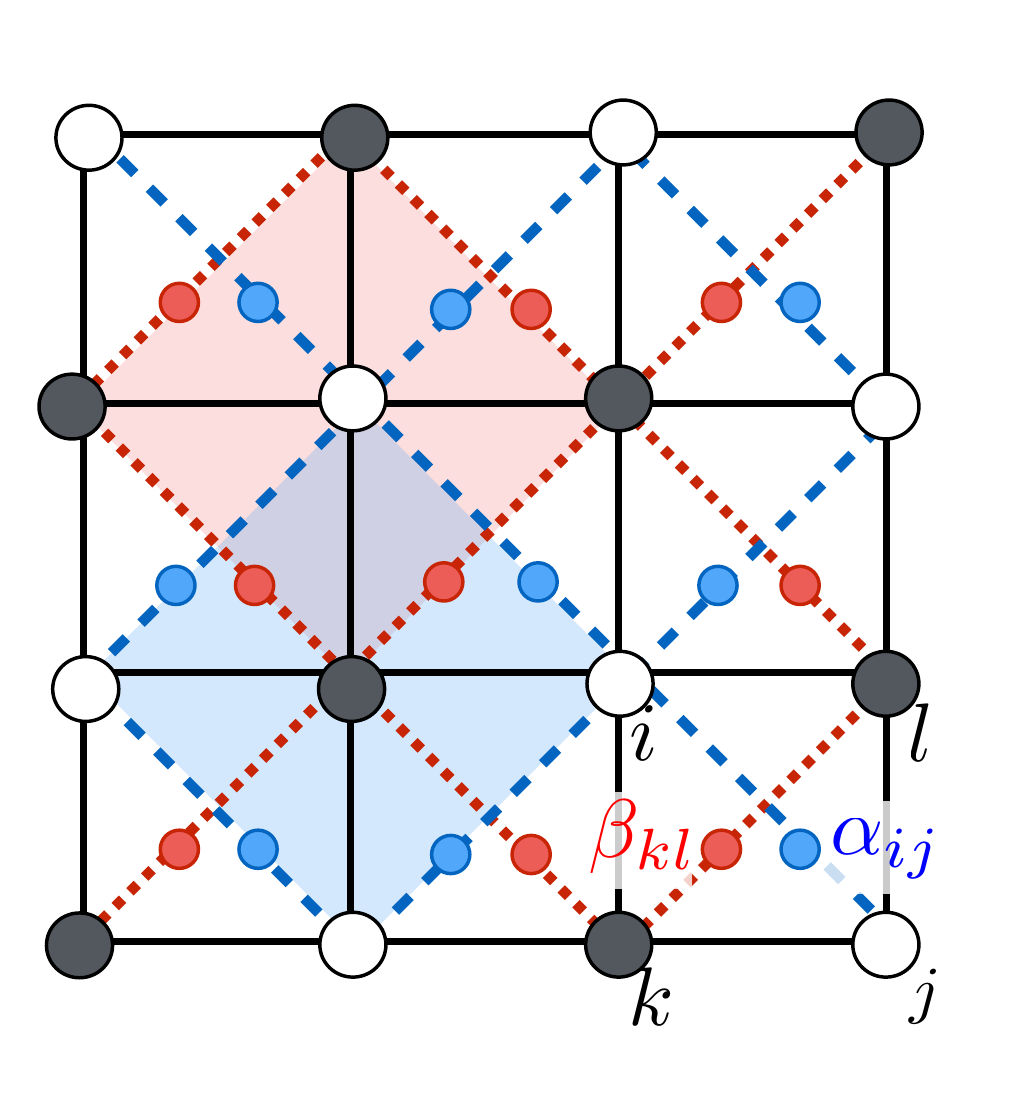}
\caption{ {\bf Gauging the FSPT model - }  The global $\Z_2^A\times \Z_2^B$ symmetry can be promoted to a local gauge symmetry, by adding $\Z_2^A$ gauge link variables $\alpha_{ij}$ (blue dots) to the bonds connecting $A$-sites (large-dashed blue lines), and $\Z_2^B$ gauge link variables, $\beta_{kl}$ (red dots) to the bonds connecting $B$-sites (small-dashed red lines). $\Z_2^{A/B}$ gauge fluxes reside on the tilted square plaquettes (shaded blue and red respectively). }
\label{fig:gauging}
\end{figure}

Let us begin with the concrete example of gauging the symmetry $G=\Z_2\times\Z_2$ protecting the dynamical 2D FSPT studied above, by coupling the model to a lattice $\Z_2\times\Z_2$ gauge field. Explicitly, let us label the factors of $\Z_2$ by their corresponding sub-lattice: $G=\Z_2^A\times\Z_2^B$. Next, we introduce dynamical Ising ``spin" variables $\alpha_{ij}^z$ and $\beta_{ij}^z$ to the diagonal links connecting the $A$ and $B$ sublattices respectively. To make the Hamiltonian in Eq.~\ref{eq:HZ2xZ2} gauge invariant, we must add gauge-link variables to the terms in $H_\text{Pump,P}$. E.g. for the plaquette shown in Fig.~\ref{fig:gauging}, we extend $\sigma^z_{1}\sigma^z_3\rightarrow \sigma^z_1\alpha^z_{13}\sigma^z_3$, and $\sigma^z_2\sigma^z_4\rightarrow \sigma^z_2\beta^z_{24}\sigma^z_4$. Furthermore, to localize the magnetic flux excitations, we add the time-independent term:
\begin{align}
H_v = -\sum_{P_A}\lambda_{P_A}\prod_{\Diamond_{P_A}} \alpha^z-\sum_{P_B}\lambda_{P_B}\prod_{\Diamond_{P_B}} \beta^z
\label{eq:gauging}
\end{align}
to Eq.~\ref{eq:HZ2xZ2}, where $\lambda_{P_{A/B}}$ are strong random coefficients, that give spatially random energies $2\lambda_{P_{A/B}}$ to $\Z_2^{A/B}$ fluxes residing on the gauge plaquette $P_{A/B}$. Finally, we must implement the gauge constraint that the product of gauge-electric field lines emanating from site $i$ is equal to the symmetry charge on that site, e.g. for the $A$ sublattice: $\prod_{j\in \<\<i\>\>}\alpha^x_{ij}=\sigma^x_i$ ($i\in A$), and similarly for the $B$ sublattice: $\prod_{j\in\<\<i\>\>}\beta^x_{ij}=\sigma^x_i$ $(i\in B)$, where $\<\<i\>\>$ denotes the next-nearest neighbors of $i$.

As a warm-up to the driven problem, let us start by recalling how to build eigenstates of the static $\Z_2\times\Z_2$ topological order obtained from the static Hamiltonian of Eqs.~\ref{eq:HZ2xZ2},\ref{eq:gauging} without the SPT pumping stages ($H_1\rightarrow 0$ in Eq.~\ref{eq:HZ2xZ2}). We may start with the topologically ordered ground-state $|\emptyset\>$ defined via $\alpha^z_{ij}, \beta^z_{ij}=+1$, and $\sigma_{A/B,i}^x=+1$. Excited states can then be created by acting with string operators that create pairs of $e_{A/B}$ and $m_{A/B}$ pairs. Explicitly, a pair of $e_A$ particles are created at positions $r,r'$ by: 
\begin{align}
W_{e_{A}}(r,r') = \sigma^x_{A,r}\sigma^x_{A,r'} \prod_{\<ij\>\in \Gamma_{rr'}} \alpha^z_{ij},
\label{eq:estring}
\end{align}
where $\Gamma_{rr'}$ is a string of edge-sharing gauge links connecting sites $r$ to $r'$, and a pair of $m_A$ particles can be created by the string operator:
\begin{align}
W_{m_{A}}(P,P') = \prod_{\<ij\>\in \Gamma^\perp_{P,P'}} \alpha^x_{ij}
\label{eq:mstring}
\end{align}
where $\Gamma^\perp_{P,P'}$ is a set of gauge links that perpendicularly intersect a string of plaquettes connecting $P$ and $P'$. There are also fermionic bound states of $\psi_{A/B}=e_{A/B}\times m_{A/B}$ created by acting with both $W_{e}$ and $W_m$ strings.

With just the static Hamiltonian, $H_2+H_v$, the eigenstates are given schematically by acting with various string operators on the ground-state, $\prod W's|\emptyset\>$, and have energy, with energy: $E = \sum_{i\in A}2h_{i}n_{e_A,i}+\sum_{i\in B} 2h_{B,i}n_{e_B,i}+\sum_{P\in A}2\lambda_{A,P}n_{m_A,P}+\sum_{P\in B}2\lambda_{B,P}n_{m_B,P}$, with respect to the vacuum, where $n_{a,r}$ are respectfully the number of quasiparticles of type $a$ at position $r$.


\subsection{Dynamical anyon permutation}
Just as for any equilibrium $\Z_2\times\Z_2$ gauge theory, particles with different $A/B$ sublattice labels have trivial mutual statistics, and distinct particles with the same $A/B$ sublattice labels are all mutual semions (exchange phase = $(-1)$). However, we expect that the symmetry properties of these excitations exhibit non-trivial dynamics due to the underlying dynamical FSPT order of the ungauged system. To see how this works, consider the effect of the pumping phase of the Floquet dynamics, in the presence of magnetic flux, $m_{A/B}$, excitations. Inserting a symmetry flux of $\Z_2^{A/B}$ through a loop of the 1D static SPT that is being pumped across the system, has the same effect as acting with the symmetry locally on one edge of that 1D SPT with open boundaries. Specifically, on the edge of the 1D $\Z_2\times\Z_2$ SPT, acting locally with one symmetry, $\Z_2^{A/B}$, inserts a charge of the other symmetry $\Z_2^{B/A}$ respectively. While we have deduced this property via simple topological arguments, one can also verify it explicitly using the above constructed lattice model, by noting that $U(T)\sigma^x_{A,i}U^\dagger(T)= (-1)^{N_{m,B}}\sigma^x_{A,i}$, where $(-1)^{N_{m,B}} = \prod_\Box \beta^z_{ij}$, is the gauge flux ($N_{m,B}$ being the vison number) on the $B$-site plaquette surrounding site $(A,i)$, and similarly for $A\leftrightarrow B$.
Since $\sigma^x_A$ is the parity of $e_A$ particles, i.e. $\sigma^x_A = +1$ if there is no $e_A$ particle, and $-1$ if there is an $e_A$ particle, the above equation indicates that an $e_A$ particle is added to the site $(A,i)$ when there exists a vison $m_B$ on the $B$-site plaquette.

Hence, as the 1D SPT is pumped across the 2D FSPT system, each $m_{A/B}$ charge binds an $e_{B/A}$ excitation with gauge-charge of the opposite $\Z_2$ symmetry. This dynamical anyon transmutation preserves the self- and mutual- statistics of all excitations, and hence does not alter the long distance topological properties. After two Floquet periods, the $m_{A/B}$ charges return to their original state, since the $e_{A/B}$ have $\Z_2$ charge. We can write the Floquet Hamiltonian as a product of evolution with respect to a local static MBL Hamiltonian, $H_\text{MBL}$, and a unitary operator, $\mathcal{P}$, that permutes anyons and cannot be generated by a fixed static Hamiltonian evolution:
\begin{align}
U_\text{FET}(T) = \mathcal{P}e^{-iH_\text{MBL}}
\label{eq:UFET}
\end{align}
Since two permutations return all anyons to their original type, we can fix $\mathcal{P}^2=1$.
Crucially, this pumping of $e_{A/B}$ particles onto the $m_{B/A}$ excitations cannot be altered by $T$-periodic perturbations of the dynamics unless these drive a bulk phase transition in the underlying topological order or lead to a spontaneous doubling of the time-periodicity by formation of a discrete time-crystal. Therefore, periodic driving enriches the set of distinct topologically ordered phases by enabling distinct patterns of dynamical anyon transmutation.

\subsection{Dynamically enforced topological degeneracies}
Since the dynamical permutation of anyons in an FET phase preserves braiding properties, it may appear rather innocuous. However, we will see that it has the dramatic consequence of enforcing large dynamically protected topological degeneracies associated with anyonic excitations.

To see this, we can repeat the trick employed in the bosonic FSPT analysis above, and look at the time-evolution operator for two periods, which, unlike the Floquet operator for one-period, can be written as the time-evolution under a local, static $\Z_2\times \Z_2$ topologically ordered Hamiltonian: $U_\text{FET}(2T) = e^{-2iH_\text{eff}}$. $H_\text{eff}$ has an emergent dynamical symmetry $\mathcal{P} = U_\text{FET}(T)e^{iH_\text{eff}}$, which permutes anyon types: $\mathcal{P}:m_{A/B}\leftrightarrow m_{A/B}\times e_{B/A}$. The eigenstates of $H_\text{eff}$ can be written as having a localized configuration of anyons of various types at positions $r$. Due to the strictly local nature of the excitations, this global symmetry acts as an effective local symmetry\cite{potter2016symmetry}, in that we can freely swap any $m_{A/B}$ excitation for an $m_{A/B}\times e_{B/A}$ excitation without changing the energy of the state with respect to $H_\text{eff}$ (subject to the global constraint of having zero net gauge charge). 

In other words, we can view the excitations as being superposed into a ``new" type of anyonic particle, $\sigma_{A/B}$, which has two internal states ($m_{A/B}$ and $m_{A/B}\times e_{B/A}$) that dictate its long-distance topological exchange statistics with other anyons. However, the dynamical symmetry $\mathcal{P}$ ``flips" this internal degree of freedom, such that a $\sigma$ with definite $\mathcal{P}$ has uncertain braiding properties with other anyons. From this, we see that the eigenstates of $U_\text{FET}(2T)$ with a fixed configuration of an even number, $N_{\sigma_{A/B}}$, of $\sigma_{A/B}$-type particles have degeneracy: $2^{N_{\sigma_A}+N_{\sigma_B}}$. This degeneracy is exactly what we would obtain from particles with non-trivial quantum dimension $d=2$. Moreover, we can see that fusing two $\sigma_{A}$ particles can yield either topological charge, $1$ (if the internal states of both $\sigma_A$'s were $m_A$ or both $m_A\times e_B$), or topological charge $e_A$ (if one $\sigma_A$ is $m_A$ and the second is  $m_A\times e_B$, or vice versa), and we see that there are two ways of obtaining either fusion outcome. The fusion possibilities for $\sigma_B$'s are similar. We can schematically write this via the non-Abelian fusion rules:
\begin{align}
\sigma_{A/B}\times \sigma_{A/B} &= 1+1+e_{B/A}+e_{B/A}\nonumber \\
\sigma_{A}\times\sigma_B &= m_Am_B+m_A\psi_B + \psi_A m_B + \psi_A\psi_B \nonumber \\
e_B\times \sigma_{A} &= \sigma_A \nonumber\\
e_A\times \sigma_B &= \sigma_B
\end{align}
which are consistent with the interpretation of the $\sigma$ particles having quantum dimension two, since the product of the quantum dimensions of anyons being fused on the left-hand sides must be equal to the sum of the squares of quantum dimensions of anyons appear in the direct sum on the right-hand sides.

We note that, while the $\sigma$ particles appear to superficially have been transmuted into non-Abelian particles, this is not quite the case. Namely, unlike truly non-Abelian particles, the degeneracy associated with the $\sigma_{A/B}$ particles is associated with local degrees of freedom associated with each anyon, and can be split by a local perturbation that breaks the $\mathcal{P}$ symmetry. In this sense, the eigenstates of $H_\text{eff}$ are largely like those of a static $\Z_2\times\Z_2$ topological order with a similar anyon permuting symmetry. One notable distinction from such a static case, is that, here, the dynamical $\mathcal{P}$ symmetry is emergent, and cannot be lifted in the absence of time-translation symmetry breaking. In other words, any $T$-periodic perturbation to the dynamical evolution will not be able to locally resolve any $\sigma$ particle into its Abelian, quantum dimension $1$ constituents. 

For a given fixed configuration of localized $\sigma$ particles, if we consider $U_\text{FET}(T) = \mathcal{P}e^{-iH_\text{eff}}$, then these $2^{N_{\sigma_A}+N_{\sigma_B}}$ degenerate states are split into $2$ sets of $2^{N_{\sigma_A}+N_{\sigma_B}-1}$ states with different overall $\mathcal{P}=\pm 1$ eigenvalue, such that the two groups of states have quasi-energy differing exactly by $\pi$. 

For example, we could consider create $\sigma$ particles in entangled pairs, via the entangled string operators:
\begin{align}
\tilde{W}^{s}_{m_{A/B}}&(r,r') = 
\nonumber\\
&\(1+s e^{i (h_{r}+h_{r'})}W_{e_{B/A}}(r,r')\)W_{m_{A/B}}(P_r,P_{r'})
\end{align}
labeled by a $\Z_2$ valued quantum number $s = \pm 1$, which respectively produce an even or odd quantum superposition of a pair of $m_{A/B}$ particles and a pair of $m_{A/B}\times e_{B/A}$ particles, with the $m$ particles residing on $B/A$-plaquettes $P_{r},P_{r'}$,  centered on $A/B$-sites $r,r'$ respectively, and the $e$ particles reside on sites $r,r'$. This entangled string operator produces a Floquet eigenstate with quasi-energy $\e_{m_A}(r,r',s) = 2\(\lambda_{B,P_r}+\lambda_{B,P_{r'}}\)+\(h_r+h_{r'}\)+\pi\frac{(1-s)}{2}$. Interestingly, the quasi-energy difference between $s=\pm 1$ is exactly $\pi$. This $\pi$ quasi-energy difference remains sharp, despite the fact that the disordered part of the Hamiltonian, $H_2$, in Eq.~\ref{eq:HZ2xZ2} would, by itself, give a non-quantized energy difference $2h_r$ between (not) having an $e_A$ particle at $r$ respectively. However, the energy difference of $H_2$ is erased from the Floquet eigenstates by the anyon permutation part of the dynamics.

Other instances of states of the internal degrees of freedom of localized $\sigma$ particles  exhibit long-range time crystalline order\cite{khemani2015phase,von2016phaseII,else2016floquet,von2016absolute,else2016pre,yao2016discrete}, for example the state with all $\sigma$ particles fixed to a definite state of either $m$ or $m\times e$, will oscillate dynamically with its $\mathcal{P}$ conjugated partner. Equivalently\cite{else2016floquet}, this state can be represented as a linear combination of two different Schr\"{o}dinger-cat Floquet-eigenstates, which are macroscopic superpositions of the fixed $m$ or $e\times m$ state and its $\mathcal{P}$ conjugated partner, whose quasi-energies differ by precisely $\pi$, and which each have one qubit of global entanglement (manifesting in long-range mutual information). 

At the level of the zero correlation length Floquet evolution, all such time-crystal and non-time crystal states are exactly degenerate and the fate of the system under more generic dynamics will be governed by how this degeneracy is lifted by residual quantum fluctuations. In the next section we explore possible fates of the FET order away from the idealized zero correlation length limit.

\subsection{Possible fates of the FET order}
The topological degeneracies enforced by dynamical anyon permutation, place strong constraints on the possible fate of the FET order away from the idealized zero correlation length system. Namely, the degeneracy is enforced by the non-commuting action of braiding of $\sigma$ particles and conjugation by the dynamical symmetry $\mathcal{P}$. This is loosely analogous to the situation of having a bunch of decoupled spins-1/2, $\vec{S}_i$, with a dihedral symmetry group with non-commuting generators $\vec{S}^{x,z}$. In the present situation, however, one symmetry generator is played by the dynamical $\mathcal{P}$ symmetry, and the role of the other symmetry generator is replaced by the action of braiding an $m_{B/A}$ particle around the $\sigma_{A/B}$ excitation. Directly analogous arguments to those of \cite{potter2016symmetry} for non-Abelian symmetry groups then show that it is not possible to have a fully localized state of these internal degrees of freedom that preserves the dynamical symmetry, $\mathcal{P}$.

\subsubsection{MBL Time-crystal}
Instead, the system could spontaneously break time-translation symmetry, forming a discrete, Floquet time-crystal\cite{khemani2015phase,von2016phaseII,else2016floquet,von2016absolute,else2016pre,yao2016discrete}. In this case, the edge states (whose topological protection relied on time-translation symmetry) are lost. Nevertheless, the FET order is imprinted even in the time-crystal phase of the system. In fact, the concept of FET order remains sharply defined despite the spontaneously enlarged symmetry group and absence of protected edge states. Further, the FET time-crystal phase can be sharply distinguished from a trivial non-FET time-crystal. 

Namely, consider the above described FET phase defined on a torus, in the MBL/time-crystal phase of the bulk excitations. Let us label the two cycles of the torus as $x$ and $y$. Then, we may consider the time evolution of a string operator, $W_{A,x}$ that creates an $m_A$ particle anti-particle pair, drags them around the $x$ axes of the torus and re-annihilates them, i.e. ``inserts an $m_A$ flux" through y-cycle. Explicitly, $W_{A,x} \prod_{\<ij\>\perp \Gamma_x} \alpha_{ij}^x$, for a path $\Gamma_x$ is a closed loop around the $x$ cycle of the torus. This operator commutes with the string operator, $W_{B,y} = \prod_{\<ij\>\perp \Gamma_y}\beta_{ij}^x$, that drags an $m_B$ particle around a path $\Gamma_y$ encircling the $y$-axis of the torus, due to the trivial mutual statistics between $m_A$ and $m_B$ particles. However, upon time-evolving for one Floquet period, $W_{A,x}$ binds an $e_B$ string, and then anticommutes with $W_{B,y}$, i.e.
\begin{align}
U(T)^\dagger W_{B,y}U(T)W_{A,x}U^\dagger(T) W_{B,y}  U(T)= -W_{A,x}.
\end{align}
This provides a sharp, formally measurable signature of the FET order, which occurs even in the regime spontaneous time-translation symmetry breaking regime, where there are no protected edge states. Moreover, this non-local property will not occur in any conventional (non-FET) time-crystal.

\subsubsection{Non-thermal critical point}
Another, potential possibility for very strong disorder, which occurs naturally e.g. in 1D chains of non-Abelian anyons, is that the system could form a non-localized but non-thermal quantum-critical-like state of the internal topological Hilbert space, such that the $\sigma$ particles remain localized. However, due to the local nature of the degeneracy in the FET systems, we expect such quantum critical states would occur as fine-tuned critical points rather than stable critical phases as for 1D non-Abelian anyons. While this possibility has been explored in a controlled analytic fashion in 1D, and to some extent in the ground-state of 2D systems, establishing whether or not such a critical scenario could occur in the excited state dynamics of 2D systems with non-Abelian particles (or FET systems with ``dynamically non-Abelianized" particles) remains an open question for future work.

\subsubsection{Prethermal FET}
A third possibility is that the system can simply thermalize due to the dynamically enforced degeneracies. However, at relatively strong disorder and for low density of excitations the system will behave MBL, and this thermalization will occur in three distinct pre-thermal stages. Namely, consider a localized state of anyonic excitations, such that the typical spacing between $\sigma$ excitations, $r_0$, greatly exceeds the single-particle localization length, $\xi$ (defined, e.g., as the localization length $H_\text{MBL}$ in Eq.~\ref{eq:UFET}, but without the anyon permutations, $\mathcal{P}$). In this limit, there are a few important, widely separated time-scales. 

First, since interactions within the non-local internal topological Hilbert space of the $\sigma$ excitations occur with amplitude $\Gamma_\text{top}\approx \Gamma_0e^{-r_0/\xi}$, where $\Gamma_0$ is roughly the anyonic ``hopping" amplitude, partial thermalization within the topological sub-space will take at least time: $t_\sigma \approx \Gamma_0^{-1}e^{r_0/\xi}$. In fact, at reasonably strong disorder, the partial thermalization time  within the internal topological Hilbert space will likely occur via highly collective rearrangements rather than  pair-wise interactions between $\sigma$-particles, which would give an even longer partial thermalization time-scale.

Subsequently, the partially thermalized internal degrees of freedom (DOF) of the $\sigma$ particles, may act as a bath to allow the anyons to delocalize. It requires overcoming an energy mismatch of order $W$, the typical disorder strength, to hop an anyon. By comparison, the bandwidth of the internal topological states associated with $n$ $\sigma$-particles is of order $\Lambda(n)\approx \Gamma_0e^{-r_0/\xi}\sqrt{n}$. Hence, to absorb this energy mismatch required to inelastically hop an anyon into the topological bath requires making of order $n\sim \(\frac{W}{\Gamma_\text{top}}\)^2 \approx \(\frac{W}{\Gamma_0}\)^2 e^{2r_0/\xi}$ excitations of the internal topological degrees of freedom. 

Assuming an eigenstate thermalization hypothesis (ETH) form\cite{srednicki1996thermal}, the local matrix element coupling between the positional motion of the $\sigma$-particle and the internal degrees of freedom of the surrounding $n$ $\sigma$-particles is of order: $\Gamma_0e^{-r_0/\xi}e^{-\frac{n}{2}\log s}$, where $s$ is the thermal entropy of the anyons. Then, a Fermi's golden rule estimate of the rate at which a $\sigma$-hopping can excited $\mathcal{O}(n)$ gives rate: $\gamma(n)\approx \frac{\Gamma_0^2e^{-2r_0/\xi}e^{-n\log s}}{\delta(n)}\approx \frac{\Gamma_0^2}{W}e^{-2r_0/\xi}$, where $\delta(n)\approx \Lambda(n) e^{-n\log s}\approx We^{-n\log s}$ is the many-body level spacing associated with the $n$ $\sigma$-particles' internal DOF,
%
%
corresponding to a time-scale $t_n \approx 1/\gamma(n)$. The amplitude for a virtual process that is off-shell in energy by an amount $W$ for time $t(n)$ is then $e^{-Wt(n)}$. From this, we see that the time-scale for delocalization of $\sigma$-particles via inelastic hopping mediated by the internal topological DOF occurs on a \emph{double-exponentially-long} timescale: $t_\text{inel.-hop} \approx \exp\[\(W/\Gamma_0\)^2e^{2r_0/\xi}\]$. Hence, there is an extremely wide separation of time scales between when the system partially thermalizes within the internal topological degenerate subspace associated with a fixed anyon configuration, and the much longer time scale upon which this partially thermalized subspace mediates anyon hopping and leads to delocalization of the entire system.

Moreover, since the system is isolated and periodically driven, if it thermalizes, we expect it to first pre-thermalize into an effective finite-temperature state with respect to some effectively static Hamiltonian\cite{Abanin15,else2016pre}, and then, at a much later time, heat up to a featureless infinite temperature state. In \onlinecite{Abanin15,else2016pre}, it was shown that the pre-thermal regime lasts up until time scale $t_*\approx e^{1/T|\delta H|}$, where $T$ is the driving period, $|\delta H|$ is an appropriate operator norm of the deviation of the Hamiltonian from a zero correlation length MBL one. Here, the dominant route to pre-thermalization is via the couplings among internal topological states of the $\sigma$ particles, we estimate that the internal topological Hilbert space will exhibit a pre-thermal regime up to time of order $t_\text{pre-therm.} \approx e^{\(1/T\Gamma_0\)e^{r/\xi}}$. The heating of the non-topological modes (e.g. via hopping of anyons), will again take at least $t_\text{inel.-hop}$. Then, depending on the relative scale of $t_\text{pre-therm.}$ and $t_\text{inel.-hop}$ there are two possible scenarios. For $t_\text{pre-therm.}\gg t_\text{inel.-hop}$, the full system will heat up to infinite temperature on times longer than $t_\text{pre-therm.}$. In the opposite case, $t_\text{pre-therm.}\ll t_\text{inel.-hop}$, the internal topological degrees of freedom will first pre-thermalize to effective finite temperature, and then heat to infinite temperature while the motional degrees of freedom of the anyons remain localized, until time-scale $t_\text{inel.-hop}$, where the full system begins to heat to infinite temperature.

\section{Floquet enriched Abelian topological order}
Having explored the features of Floquet enriched topological order obtained by gauging a bosonic FSPT, we now explore several generalizations. We begin with a general discussion of the structure of FET phases, building off of the concrete example of the Floquet enriched $\Z_2\times\Z_2$ topological order analyzed in the previous section. Then, we discuss several concrete examples, beginning with gauged-versions of other bosonic FSPTs, followed by examples of other ``beyond-cohomology" FET phases that cannot be obtained by a gauging a bosonic FSPT, but rather arise from dynamical pumping of emergent non-bosonic anyon degrees of freedom.

\subsection{General structure of FET phases}
For the gauged bosonic FSPT case above, we saw that the Floquet evolution of this FET phase was characterized by a dynamical permutation of anyon types, which preserved the topological structure of the permuted anyons. More dramatically, some initially Abelian gauge-flux particles were dynamically transmuted into effectively non-Abelian particles. In this section, we describe the structure of generic Abelian FET phases in general terms, and then consider several concrete examples beyond the Floquet enriched $\Z_2\times\Z_2$. 

In all the examples we will encounter, the Abelian FET phases start with a topological order characterized a set of anyons $\{1,a,b,c,\dots\}$, in which the driving implements non-trivial statistics preserving dynamical permutations of the anyon types: $a_1\overset{\tiny U(T)}{\longrightarrow} a_2\overset{\tiny U(T)}{\longrightarrow} \dots a_1$, with anyon $a$ cycling back to itself after $N_a$ periods. For example, in the $\Z_2\times\Z_2$ FET obtained from gauging the bosonic FSPT lattice model above, we have: $U(T):m_A\rightarrow m_A\times e_B\rightarrow m_A$, and similarly for $A\leftrightarrow B$, i.e. all anyons return to themselves after at most two Floquet periods ($N_{m_A}=2$, etc...). 


In general, we can write the time-evolution operator for such an FET phase as:
\begin{align}
U(T) = \mathcal{P}e^{-iH_\text{MBL}}
\end{align}
where $\mathcal{P}$ permutes the anyon types and satisfies $\mathcal{P}^{\text{LCM}\(\{N_a,N_b,\dots\}\)} =1$, where $\text{LCM}$ denotes the least common multiple, and $H_\text{MBL}$ is a topologically ordered MBL Hamiltonian, that gives (random) on-site energy $E_a(r)$ for having an anyon of type $a$ at position $r$ (or more generically is related to this idealized case by a finite depth local unitary deformation).

In the zero-correlation length limit, a generic eigenstate of the static (non-FET) Abelian topological order, would consist of starting with the topological vacuum (ground-state), $|\emptyset\>$, and acting with string operators of the form: $W_a(r,r')$ that create an anyon of type $a$ at $r$ and its anti-particle $\bar{a}$ at position $r'$: $|\Psi_\text{static}\> = \prod_I W_{a^I}(r_I,r'_I)|\emptyset\>$. However, FET eigenstates must clearly be a superposition of such static states with different anyon types, since time evolution permutes the anyon types:
\begin{align}
U(T) W_{a_{1}}(r,r')U^\dagger(T) = e^{-i\(E_{a_1}(r)+E_{\bar{a}_1}(r')\)}W_{a_{2}}(r,r')
\end{align}
where $E_{a}(r)$ is the contribution to the quasi-energy from having an anyon type $a$ at position $r$. 

However, we can construct entangled superpositions of the string operators, which create Floquet eigenstates:
\begin{align}
\tilde{W}^{j}_{[a]}(r,r')\equiv \frac{1}{\sqrt{N_a}}
\sum_{j=1}^{N_{a_0}} e^{2\pi i j/N_a} e^{-i\(E_{a,j}(r)+E_{\bar{a},j}(r')\)}W_{a_j}(r,r')
\label{eq:entangledstring}
\end{align}
where, $[a]$ denotes the equivalence class of anyons related to $a$ by some number of time-evolution periods, $E_{a,j}(r) = \sum_{k=1}^{j-1}E_{a_k}(r) -\bar{E}_a(r)$, and similarly $E_{\bar{a},j}(r') = \sum_{k=1}^{j-1}E_{\bar a_k}(r')-\bar{E}_{\bar a}(r')$, and $\bar{E}_a(r) = \frac{1}{N_a}\sum_{j=1}^{N_a}E_a$, is the on-site energy of anyons $a_j$ averaged over the cycle of dynamical permutations. One can readily verify that the operators $W^{j}_a(r,r')$ create Floquet eigenstates with quasi-energy: $\bar{E}_a(r)+\bar{E}_a(r')+\frac{2\pi j}{N_a}$. In other words, there are $N_a$ different superpositions of a given string operator labeled by $j=1,\dots N_a$, whose quasi-energies differ by exactly an integer multiple of $\frac{2\pi}{N_a}$. 

As in the above example, we can introduce a generalized type of excitation, denoted $[a]$, which can take the form of any of the Abelian anyons $a_{1\dots N_a}$, or quantum superpositions of these Abelian particles. We can associate an internal $\mathbb{Z}_{N_a}$-valued degree of freedom with each $[a]$ particles that takes definite value when $[a]$ is in a definite state of being a fixed Abelian anyon from the possible set of $a_{1\dots N_a}$ particles. This internal degree of freedom is incremented by the dynamical symmetry $\mathcal{P}$, such that $[a]$ particles with definite Abelian exchange statistics are not eigenstates of $\mathcal{P}$ and vice versa. From here, we can immediately see that there is a dynamically enforced topological degeneracy of Floquet eigenstates that scales as $\sim N_a^\text{\#~[a]~particles}$, indicating that the $[a]$ particles have effective quantum dimension $d_{[a]}=N_a$.

The $[a]$ particles also exhibit non-Abelian fusion properties, expressed via the fusion rules:
\begin{align}
[a]\times[b] = \sum_{i=1}^{N_a}\sum_{j=1}^{N_b} a_i\times b_j.
\end{align}

 We note that the effective non-Abelian properties of the dynamically permuted anyons are closely related to those of a static system with a symmetry that permutes anyon types \cite{barkeshli2014symmetry,khan2014anyonic,teo2015theory,tarantino2016symmetry}, which upon gauging of this global symmetry generically produces a non-Abelian topological order\cite{barkeshli2014symmetry,teo2015theory}. 
However, we remark that the dynamical anyon permutation symmetry is ``emergent" and stable to any local $T$-periodic perturbation, and hence can only be lifted by a local perturbation with an enlarged $N_aT$ period. 

Having explored the idealized regime of zero-correlation length, let us consider more generic models obtained by adding generic $T$-periodic perturbations. This induces longer-range interactions among anyons that decay exponentially in the separation of the two anyons measured in units of the localization length. As discussed in the previous section for the particular case of a gauged bosonic FSPT, the system cannot remain in a time-translationally symmetric MBL state due to the effective non-Abelian properties of the dynamically permuted anyon excitations. Rather, at strong enough disorder to stay in the MBL regime, the FET order is necessarily accompanied by spontaneous breaking of time-translation symmetry with an effective $N$-tupling of the fundamental period, with $N=\text{LCM}\(\{N_a,N_b,\dots\}\)$.

\subsection{FETs from other gauged boson FSPTs}
Various other related examples of dynamical FET phases, can be obtained by gauging the symmetries of other 2D FSPTs with discrete gauge groups. 
To obtain the general structure for discrete Abelian gauge groups, it is sufficient to consider topological order with a discrete Abelian gauge group of the form: $G=\Z_{N_A}\times\Z_{N_B}$.

Distinct FET phases of the $\Z_N\times\Z_M$ topological order are characterized by a dynamical pumping of $j_A$-units of $A$-electric charge to $B$-fluxes, and $j_B$-units of $B$-electric charge to $A$-fluxes, where $j_{A}\in \Z_{N_A}$ and $j_B\in \Z_{N_B}$. However, not all $j_{A/B}$ are allowed. Rather, since the pumping occurs via local dynamics, it cannot change the long-range topological braiding properties of the underlying anyons. This requires: $N_Bj_A = -N_A j_B$, constraining $j_A =\frac{N_A \ell}{\text{gcd}(N_A,N_B)}$ and $j_B =\frac{N_B \ell}{\text{gcd}(N_A,N_B)}$, where $\ell\in \{1\dots \text{gcd}(N_A,N_B)\}$, and $\text{gcd}$ denotes the greatest common denominator. These FET phases are again in one-to-one correspondence with bosonic 1D static SPTs protected by $\Z_n\times\Z_m$ symmetry, which can be pictured as condensates of $\Z_{N_A}$ domain walls bound to $j_B$ charges of $\Z_{N_B}$ and of $\Z_{N_B}$ domain walls bound to $j_A$ charges of $\Z_{N_B}$, with integers $j_{A/B}$ subject to the same $N_Bj_A=-N_Aj_B$ constraint as described above. Again, we can construct explicit solvable lattice models of these FET phases by coupling the FSPT models constructed in App.~\ref{app:BosonModels}, to a $\Z_{N_A}$ and $\Z_{N_B}$ valued gauge links as shown in Fig.~\ref{fig:gauging}.

The dynamical permutation of these phases, changes $m_A$ into a quantum superposition, $\chi_A$, of $m_A\times e_B^{\ell j_B}$ for all $\ell = \{0,1,\dots (N_B\hspace{2pt}\text{mod}\hspace{2pt} j_B)-1\}$, thus dynamically turning $m_A$ into an effectively non-Abelian object $\chi_A$ with dimension $d_{\chi_A}=N_B\hspace{2pt}\text{mod}\hspace{2pt} j_B$, and fusion rules:
\begin{align}
\chi_A\times\chi_A = 2\sum_{\ell =0}^{d_{\chi_A}-1}\(m_A\times e_B^{\ell \cdot j_B}\).
\end{align}
Similarly, $m_B$ is dynamically superposed into an effectively non-Abelian object $\chi_B$, with quantum dimension and fusion rules obtained from the above expressions with $A\leftrightarrow B$.

\subsection{Fermionic Floquet enriched topological orders}
So far, we have encountered many examples of 2D bosonic FSPTs, and found that gauging the static symmetry group of a 2D bosonic FSPT, results in a dynamical topological order enriched by periodic driving. What about corresponding Floquet topological phases of fermionic matter? 

To start, let us restrict our attention to the classes of symmetries described in the 10-fold way classification, namely those with a subset of charge-conservation, particle-hole, and time-reversal symmetries. Fermionic SPT phases with these symmetries face fundamental problems that prevent them from being localized, as explained in \cite{potter2016symmetry,potter2016topological}, and hence will be unstable to heating. To summarize these arguments: phases with conserved charge require particle-hole symmetry to be non-trivial, which makes the symmetry group non-Abelian and fundamentally prevents localization. Fermion SPT phases without charge-conservation face a less fundamental but equally serious obstacle that rules out their physical realization: namely, MBL is spoiled by Coulomb interactions, and hence must occur in systems of charge-neutral fermions (e.g. neutral fermionic atoms). Here, number conservation symmetry can only be broken spontaneously, resulting in a superfluid phase with a gapless goldstone mode which acts as a thermalizing bath that destroys MBL (much like phonons in a solid).

In contrast, neutral fermionic excitations may occur as emergent anyonic particles in topologically ordered systems. If these fermionic excitations form Floquet SPT states, the resulting topologically order will exhibit non-trivial dynamics giving rise to new examples of FET phases. To explore these phases, we will start by examining potential 2D fermion FSPTs in systems where the fermion charge is not conserved, with an eye towards coupling the fermions to a fluctuating $\Z_2$ gauge field to obtain fermionic FET phases.

\subsubsection{Fermionic Floquet enriched $\Z_2$ topological order}
\begin{figure}[t!]
\includegraphics[width=\columnwidth]{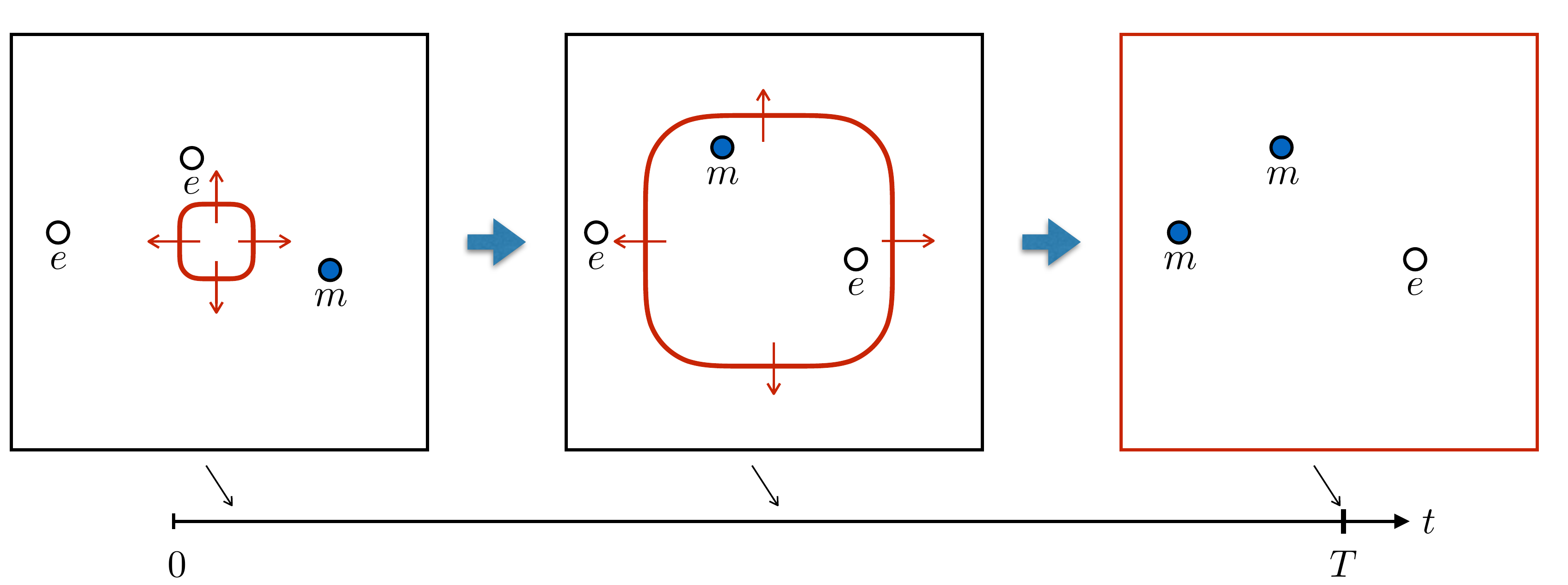}
\caption{ {\bf Dynamical anyon permutation - } Schematic depiction of dynamical anyon permutation in a Floquet enriched $\Z_2$ gauge theory, in which a topological superconducting chain (rounded red rectangle with arrows) of the $\psi = e\times m$ particles is pumped from the bulk onto the boundary during each period. Anyonic $e$ and $m$ excitations are permuted as they pass through the topological chain. }
\label{fig:anyonperm}
\end{figure}

In the absence of any symmetry, there is a single non-trivial 1D fermion SPT: a topological superconducting chain with unpaired Majorana fermion edge states. This enables a 2D FSPT phase in which the 1D fermion SPT is dynamically pumped onto the boundary during every Floquet period. Gauging the fermions of this 2D fermionic FSPT phase, results in a FET $\Z_2$ gauge theory with mutual semionic excitations: $\{1,e,m,\psi\}$, where $\psi$ is the gauged fermion, and $e$ and $m$ are bosonic $\Z_2$ gauge charge and flux excitations. The 1D fermion SPT pumping has the effect of interchanging the $e$ and $m$ particles each Floquet period as illustrated in Fig.~\ref{fig:anyonperm}. To see this, note that inserting $\pi$ flux into a loop of the 1D topological superconductor flips the fermion parity of that loop. Since the $\psi$ fermions see the $e$ and $m$ particles as $\pi$ fluxes, the 1D topological superconducting loops flip fermion parity when dragged over either an $e$ or an $m$ particle. Since the total fermion parity must be conserved, a corresponding $\psi$ fermion must be fused onto the $e$ or $m$ particles, interchanging them (since $e\times \psi = m$ and $m\times \psi = e$).

Again, we see that the FET phase differs from an ordinary topologically ordered phase with the same topological properties, in that the anyons are dynamically permuted by the periodic driving in a way that preserves their self- and mutual- braiding statistics. However, this FET phase is missed by considering gauged boson FSPT phases, for which there are no non-trivial phases protected by only $\Z_2$ symmetry.

As for the bosonic FSPT phases described above, the dynamical pumping tunes the edge to a self-dual point equally poised between the topological superconductor and trivial superconducting phases. We note that, since it is not possible to break the gauge symmetry, the edge of this FSPT phase cannot be trivially localized even at very strong disorder. The only options are a thermal edge, a time-crystal edge with spontaneously doubled periodicity, or a quantum critical edge that preserves symmetry.

Further, the $e$ and $m$ particles become dynamically quantum superposed into an effectively non-Abelian particle $\sigma$, with effective quantum dimension $2$, and fusion rules:
\begin{align}
\sigma\times\sigma &= 1+1+\psi+\psi \nonumber\\
e\times\sigma &= m\times \sigma = \sigma.
\label{eq:IsingFusion}
\end{align}

\subsubsection{Fermionic Floquet enriched $\Z_2\times\Z_2$ topological order }
By gauging the symmetry group of bosonic FSPT phases with $\Z_2$ symmetry we encountered a new FET phase. However, we just saw that such gauged bosonic FSPT phases do not exhaust the possible FET phases. Are there new FET phases with this topological order can be obtained by gauging the fermion parity and on-site symmetry of a 2D fermions SPT with $\Z_2$ symmetry? 

Our strategy so far has been to obtain an FET by dynamically pumping 1D SPTs. There are four types of intrinsically fermionic 1D SPTs with $\Z_2$ symmetry characterized by two $\Z_2$-valued topological indices: $\nu_{A/B}\in \{0,1\}$, corresponding to the number of $\Z_2$ charged (A) and neutral (B) Majorana fermion end states. 

Naively one might suspect that each of these 1D phases produces a distinct FET phase characterized by the dynamical anyon permutations:
\begin{align}
e_\pm \rightarrow \begin{cases} e_\pm &; \nu_\pm = 0 \\ m_\pm &; \nu_\pm = 1 \end{cases}
\end{align}
where we have labeled the $e$ and $m$ particles for each $\Z_2$ sector by a $\pm$ label. These permutations appear superficially different than that of the bosonic FSPT: $m_\pm \rightarrow e_\mp m_\pm$. However, the case with $\nu_+=\nu_-=1$ is actually equivalent to the bosonic one up to a trivial relabeling of anyons: $\tilde{e}_\pm \equiv e_\pm$, $\tilde{m}_\pm \equiv e_\mp m_\pm$. On the other hand, the phases with only one of $\nu_\pm$ nontrivial have dynamical anyon permutations only in one or the other $\Z_2$ gauge sector, and are equivalent to each other up to relabeling of anyons, but distinct from the bosonic one. Hence we see that there are only two distinct non-trivial FET phases. Again, the dynamical permutation of $e_\pm$ and $m_\pm$ particles, promotes these objects into quantum dimension $2$ particles, $\sigma_{\pm}$, with fusion rules identical to Eq.~\ref{eq:IsingFusion}.

\subsection{Floquet enriched $\Z_n$ topological order -- pumping parafermion chains}
The general pattern has been that the FET phase is characterized by a dynamical permutation of anyons that leaves the braiding statistics invariant. One may then suspect that any such topology-preserving dynamical anyon permutation can be realized in a an appropriate FET phase. For example, consider a $\Z_N$ gauge theory, containing bosonic $\Z_N$ charges, $e$, and fluxes $m$ ($e^N=1=m^N$), and their composite particles. Since the topological properties of this $\Z_N$ gauge theory are invariant under interchanging $e\leftrightarrow m$, one might guess that there is a possible FET phase in which this permutation occurs dynamically every Floquet period. 

So far all of the examples of FETs we have encountered can be viewed as gauged bosonic or fermionic FSPTs. However, in this case, for $N>2$, one can readily verify that there are no such bosonic or fermionic FSPTs, that produce the $e\leftrightarrow m$ upon gauging. Is the above-described dynamical anyon permutation possible in a Floquet enhanced $\Z_N$ topological order?

\subsubsection{Topological chains of anyons}
To shed some light on this question, note that in the above fermionic examples, we constructed a 1D SPT of an emergent particle. While in the previous cases, this emergent particle was a familiar fermion, generically we can attempt to make 1D ``SPT" phases out of any emergent anyonic excitation -- not just those that are fermionic or bosonic. Such phases would not be possible in a truly 1D system, since in our universe there are only fermions and bosons microscopically. However, a 2D topological order has no such preconceived prejudices. In fact, we will see that the $e\leftrightarrow m$ permuting FET phase can actually be viewed as being induced by dynamical pumping of 1D FSPT chains of a particular emergent anyon.

Specifically, in the $\Z_N$ topological order, consider the bound state, $\psi = e\times m^{-1}$ of the $\Z_N$ charge and flux. The mutual statistics between charge and flux is: $\theta_{e,m} = e^{2\pi i/N}$, making their bound state $\psi$ an anyonic particle with statistics $e^{-2\pi i/N}$ that is neither fermionic nor bosonic for $N>2$. We can construct a 1D array within this 2D $\Z_N$ gauge theory, where each site, $i=1\dots L$, contains an $N$-state Hilbert space labelled by the number of $\psi$ particles: $n_i\in \Z_N$, or more conveniently by their charge: $Q_i = e^{2\pi i n_i/N}$. The total Hilbert space of the chain is subject to a global constraint that the total number of $\psi$ particles is $0\mod N$ since the total system must be gauge neutral, (this generalizes the constraint of having even number of fermions for $N=2$). 

Following \cite{fendley2012parafermionic,motruk2013topological} in an analogous route to Kitaev's construction of the topological superconducting chain, we may introduce a pair of $\Z_N$ parafermion operators: $\chi_{2i},\chi_{2i+1}$ for each site of the anyon chain. These operators satisfy: $(\chi_i)^N=1$, $\chi_i\chi_j = e^{(2\pi i/N)\Theta(i-j)}\chi_j\chi_i$, and $\chi_i^\dagger = \chi^{N-1}_i$, where the $\Theta(x)$ is a step function. The total $\psi$ charge on site $i$ is measured by $Q_i = (-e^{-i\pi/N})\chi_{2i}\chi_{2i+1}$. From this and the commutation relations, we see that $\chi_{2i}^\dagger$ and $\chi_{2i+1}$ both destroy a $\psi$-particle on site $i$. Crucially, the nontrivial commutation relations among these parafermionic operators ensures the appropriate commutation relations among the $\psi$ particles. Moreover, while the parafermionic degrees of freedom have fractional quantum dimension $d_\chi = \sqrt{N}$, they are constrained to appear in pairs so that the Hilbert space of each site agrees with that of the $\psi$-particles, so that there are no unphysical non-Abelian anyonic excitations in this description.

As was derived in \cite{fendley2012parafermionic,motruk2013topological,clarke2013exotic,lindner2012fractionalizing}, a parafermionic generalization of Kitaev's chain is obtained by the Hamiltonian: 
\begin{align}
H_\text{PFC} = -t\sum_{i=0}^{L-2} e^{i\pi/N}\chi^\dagger_{2i+1}\chi_{2i+2}+\text{h.c.},
\label{eq:hpfc}
\end{align}
which, in an open chain leaves free parafermions at the edge of the chain. Similarly, generalizing the topological superconductor, adding $f$-units of $\Z_N$ flux to the center of a loop of topological parafermion chain, corresponds to changing the value of one of the $\psi$-charge of one of the bond operators $\chi_{2i+1}^\dagger\chi_{2i+2}$ (which one depends on gauge) from its ground-state value of $1$ to $e^{2\pi if/N}$ terms of the ring. This, in turn, corresponding to adding $\psi$-charge to the chain. 

\subsubsection{FET phase from pumping parafermion chains}
Then, we can envision an FET phase obtained by pumping these 1D topological parafermion chains onto the boundary of the system during each Floquet period, analogous to the 2D FSPT cases described in detail in previous sections. Again repeating the arguments of previous sections, we see that an $e$ particle will appear as a $+1$ $\Z_N$ flux to the $\psi$ particles, inducing a $\psi$ charge on the topological loop being pumped, and a compensating $\psi^{-1} = e^{-1}\times m$ will bind to the $e$, transmuting it to an $m$ particle (similarly the $m$ will appear as $(-1)$ flux for the parafermion chains, and will be transmuted into an $m$ particle).

As for previous examples, the dynamical interchange of $e\leftrightarrow m$ particles in this this Floquet enriched $\Z_N$ gauge theory, effectively turns these Abelian particles into effectively non-Abelian objects $\chi$ which can be any quantum superposition of anyon types $e\times \psi^{j}$ for $j=0$ to $N-1$. The $\chi$ particles have effective quantum dimension $N$, and obey fusion rules:
\begin{align}
\chi\times\chi &= 2\sum_{j=0}^{N-1}\psi^j\nonumber\\
\psi^j\times \chi &= e\times \chi = \chi.
\end{align}
%


\subsection{Floquet enriched non-Abelian topological orders}
So far we have restricted our attention to FET phases arising from driving Abelian topological orders. One could certainly envision an even richer set of phases obtained by driving non-Abelian topological orders. Here, even the un-driven system cannot fully obtain a fully MBL state\cite{potter2016symmetry,vasseur2015quantum}. However, these orders could arise in an exponentially long-lived metastable pre-thermal regime\cite{Abanin15,else2016pre}, or possibly as quantum critical phases that are neither MBL nor thermal\cite{vasseur2015quantum}. Likely, one could also obtain phases in which there is a dynamical anyon permutation symmetry. The possible non-anomalous set of such anyon permutation symmetries in non-Abelian phases have been classified in \cite{barkeshli2014symmetry}. Understanding whether all of these can be obtained by Floquet driving remains a challenge for future work.

\header{Discussion}
In this paper, we have constructed exactly solvable models for 2D FSPT phases. We showed that these phases can be understood as undergoing a dynamical pumping of a lower dimensional static SPT onto the boundary during each Floquet cycle, which we showed tunes the edge to a self-dual Hamiltonian, which cannot be trivially localized while preserving symmetry. We derived a conceptual diagnostic for the dynamical anomaly protecting the edge in terms of symmetry flux insertion. Interestingly, this anomaly is visible even when the edge is in an infinite temperature thermal state. 

We then explored how dynamical driving can enrich intrinsic topological order. A subset of such FET states were obtained by promoting the global symmetry of a 2D bosonic FSPT to a local gauge symmetry. Such phases can be classified by cohomology formulation by enlarging the gauge group to include an extra integer factor of $\Z$ corresponding to time-translation, in analogy to equilibrium Dijkgraaf-Witten theories. However, we found that these cohomology states represent only a fraction of the possible FET states. We construct a variety of ``beyond cohomology" FET states, which can be viewed as 2D topological ordered states, in which 1D topological chains of emergent anyons (not necessarily bosons or fermions) are dynamically pumped across the system in each driving cycle. Equivalently, we showed that this pumping has the effect of dynamically permuting the anyonic excitations in a way that preserves their braiding statistics. These results, we conjectured that 2D FETs are generally classified by exchange-statistics-preserving anyon permutations (modulo trivial relabeling of anyon types). However, while we are not aware of any counterexamples, we are currently unable to rule out the possibility of certain anyon permutations are anomalous and not realizable in a local 2D Floquet-MBL system.

While we have built a solid understanding of many new non-equilibrium 2D dynamical topological phases, many open questions remain. For example, 3D SPTs systems can exhibit exotic surface topological orders with anomalous symmetry properties that make them impossible to realize in a purely 2D system. This feature presumably extends to 3D FSPTs. Understanding this resulting anomalous surface FET (or more properly Floquet and symmetry enriched topological order, or FSET) remains an open challenge, and how it differs from two dimension examples remains an open challenge. Furthermore, 3D topological order is only partially understood in equilibrium. Extending these concepts to driven non-equilibrium settings presents even further challenges.

We close with a note of caution. Throughout this work, we have assumed that MBL is stable in higher dimensions. We note, however, that there has recently been some work casting doubt on the stability of higher dimensional MBL systems to thermalization by exponentially rare thermal regions\cite{deroeck2016stability}. In our view, these arguments are not conclusive, and the issue remains unsettled. However, in light of these doubts, a pessimistic reader can be reassured that our results represent, at worst, an accurate description of the dynamics of the systems up to parametrically long time scale that diverges exponentially with disorder strength. In a practical sense, experimental realizations are anyway inevitably cut short at finite times due to imperfect isolation from the environment, and the rare thermal regions are unlikely to be a limiting factor for strong disorder.

\vspace{4pt}\noindent{\bf Acknowledgements -- } We thank R. Vasseur, H.C. Po, and especially A. Vishwanath for insightful conversations. This work was supported by University of Texas at Austin startup funds (ACP), and the Gordon and Betty Moore FoundationÕs EPiQS Initiative Theory Center Grant (TM).

\appendix

\section{Explicit model for 2D FSPT with $\Z_n\times\Z_n$ symmetry \label{app:BosonModels}}
In this section, we generalize the model above constructed for 2D FSPTs with $\Z_2\times\Z_2$ symmetry, to build analogous models for $\Z_n\times\Z_n$ symmetric 2D FSPTs, which form the basic building block for general 2D bosonic FSPTs (i.e. those with finite-Abelian symmetry groups, as required to have many-body localization to obtain a well defined Floquet phase). The models are closely related to the one analyzed in the main text. Namely, consider a 2D lattice of sites with $n^2$ different states, which we can write as a square lattice of $n$-state spins with a $A$ and $B$ sub-lattice labels, so that each physical site contains both an $A$ and $B$ spin. We introduce the $\Z_n$ generalizations of the Ising-spin operators:
\begin{align}
Z_i = \begin{pmatrix} 
1 & &&& \\ 
&\omega &&& \\
&& \omega^2 && \\ 
&&&\ddots& \\ 
&&&&\omega^{n-1} \end{pmatrix} 
,\hspace{0.1in}
\omega \equiv e^{2\pi i/n}
\label{eq:Zn}
\end{align}
as well as the cyclical spin ``raising" operators:
\begin{align}
X_i = \begin{pmatrix} 
0&1&0&\cdots&0 \\ 
0&0&1&\cdots&0 \\
\vdots & \vdots&\vdots &\ddots& \vdots \\
0&0&0&\cdots&1 \\ 
1&0&0&\cdots&0 \end{pmatrix}
\end{align}
that satisfy $X_i^n = 1 = Z_i^n$, and $X_i^\dagger Z_i X_i = \omega Z_i$. We will denote the symmetry group as $\Z_n^A\times\Z_n^B$, so that we may refer to the different factors of $\Z_n^{A/B}$ without ambiguity, and consider the symmetry generators to act as:
\begin{align}
g_A& = \prod_{i} Z_{2i} \nonumber\\
g_B&= \prod_{i} Z_{2i+1}
\end{align}

We again aim to build a Floquet drive that pumps a 1D SPT onto the boundary for each cycle. To this end, we begin by constructing zero correlation length models of the 1D SPTs protected by $\Z_n^A\times\Z_n^B$ symmetry. There are $n$ distinct such phases, which can be viewed as condensates of $\Z_n^A$ domain walls bound to $j$-charge of $\Z_n^B$ of symmetry. Consistency requires that $\Z_n^B$ domain walls bind $(-j)$-charges of $\Z_n^B$ symmetry. An explicit zero-correlation length model that implements this construction is:
\begin{align}
H_\text{1D} = \sum_{i} \lambda_{2i} Z_{2i-1}^\dagger X_{2i}Z_{2i+1} +\lambda_{2i+1} Z_{2i}X_{2i+1}Z_{2i+2}^\dagger +h.c.
\end{align}
where $\lambda_i$ are spatially random complex coefficients. On an open chain of length $L$, $1<i<L$, this model contains free $\Z_n$ spins on each edge. E.g. on the left edge, the operators $\mathcal{Z}_L\zeta\equiv Z_1$ and $\mathcal{X}_L\equiv X_1Z_2^\dagger$ commute with the Hamiltonian, and, having the same commutation relations as $Z_1$ and $X_1$, form an effective $\Z_n$-spin. However, the symmetries act this effective edge spin act as $(g_B)_\text{L,edge} = \mathcal{Z}_L$, $(g_A)_\text{L,edge} = \mathcal{X}_L$, which form a projective representation with $(g_Ag_Bg_A^{-1}g_B^{-1})_\text{L,edge} = \omega^{-1}$.

To build a 2D FSPT, we then seek a unitary operator that adds a 1D SPT phase onto a loop of sites, i.e. which takes:
\begin{align}
U_\text{SPT}:\begin{cases}X_{2i}~~\rightarrow &Z_{2i-1}^\dagger X_{2i}Z_{2i+1} \\
X_{2i+1}\rightarrow &Z_{2i} X_{2i+1}Z_{2i+2}^\dagger 
\end{cases}
\label{eq:USPTZn}
\end{align}
The appropriate operator can be identified by taking a temporary conceptual detour, in which we consider the $\Z_n$ spin to be embedded in a continuous $U(1)$ rotor variables: $X_i\rightarrow e^{i\phi_i}$, where $\phi_i\in[0,2\pi)$, and then after having identified the correct unitary transformation in the rotor language, we can restrict back to the discrete $\Z_n$ configurations, $\phi_i \in \left\{\frac{2\pi j}{n}\right\}_{j=0}^{n-1}$. In this continuous $U(1)$ representation, $Z_i$ acts like $\omega^{\ell_i}$, where $\ell_i$ is the momentum conjugate to $\phi_i$. Then, the operator $e^{i\alpha\ell_i}$ acts like a translation operator that shifts $\phi_i$: e.g. $e^{-i\alpha\ell_i}e^{i\phi_i}e^{i\alpha \ell_i}=e^{i\alpha}e^{i\phi_1}$. Then, we see that the rotor operator that implements the desired transformation in Eq.~\ref{eq:USPTZn} for sites in a closed loop, $\Gamma$, is:
\begin{align}
\(U_{\text{SPT},\Gamma}\)_\text{rotor} = \prod_{j\in \Gamma}\omega^{(-1)^j\ell_j\ell_{j+1}}
\end{align}
Since this operator always shifts $\phi_i$ by an integer multiple of $2\pi/n$, we can then simply restrict it to the rotor basis. This $U_\text{SPT}$ operator, then commutes with symmetry, and can be written as $e^{-H_{\text{pump},\Gamma}}$, where $H_{\text{pump},\Gamma}$ involves multiple spin operators of size up to the length of $\Gamma$.

Just as for the $n=2$ case described in the main text (Eq.~\ref{eq:HZ2xZ2}), we can then use these ingredients to write down a model for the 2D $\Z_n\times\Z_n$ root FSPT which pumps the minimal 1D SPT onto the edge during each cycle:
\begin{align}
H(t) = 
\begin{cases}
2H_1 = 2\sum_P H_{\text{pump},P} & 0\leq t< 1/2
\vspace{4pt} \\ 
2H_2 = \sum_i h_i^{(j)} (X_i)^j +h.c.) & 1/2\leq t<1
\end{cases}
\label{eq:HZnxZn}
\end{align}
where $h_i^{(j)}$ are random complex coefficients, such that $U(T)=e^{-iH_2}U_{\text{SPT},\d\Omega}$, where $\d\Omega$ is the boundary of the system.

Again, we may consider the edge theory resulting from this Hamiltonian. It is now convenient to consider the Floquet evolution operator for $n$ periods:
\begin{align}
U(nT)_\text{edge} = \(e^{-iH_2}U_{\text{SPT}}\)^n = \prod_{j=0}^{n-1} e^{-i(U_\text{SPT}^\dagger)^j H_2(U_\text{SPT})^j}
\end{align}

\begin{widetext}
We see that the edge is constantly cycled among the different $n$ SPT 1D SPT phases. We can analyze the resulting edge dynamics explicitly in the limit of $|h_i^{(j)}|\ll 1$. In this limit, defining $U(nT)=e^{-iH_\text{eff}^{(n)}}$, we have:
\begin{align}
H_\text{eff}^{(n)}\approx \sum_{j,k=0}^{n-1}\sum_i \[h_{2i}^{(k)}\(Z_{2i-1}^\dagger\)^{jk} X_{2i}^k \(Z_{2i+1}\)^{jk}+h_{2i}^{(j)}\(Z_{2i-1}\)^{jk} X_{2i}^k \(Z_{2i+1}^\dagger\)^{jk}\]+h.c.+\dots
\end{align} 
where $\dots$ are subleading in $|h_i^{(j)}|$, but preserve the cyclic ``symmetry" associated with the edge Hamiltonian being invariant under incrementing the 1D SPT invariant of the edge ($j\rightarrow j+1$).

As for the $n=2$ case, it is instructive to map this Hamiltonian into two $\Z_n$ clock models, via the duality transformation:
\begin{align}
Z_{2i-1}^\dagger X_{2i}Z_{2i+1}&\rightarrow \zeta_{A,i}\zeta_{A,i+1} \nonumber \\
Z_{2i}X_{2i+1}Z^\dagger_{2i+2}&\rightarrow \zeta_{B,i}\zeta_{B,i+1}\nonumber \\
X_{2i-1}&\rightarrow \xi_{A,i}\nonumber \\
X_{2i}&\rightarrow \xi_{B,i}
\end{align}
where the dual variables $\zeta_{A/B}$ and $\xi_{A/B}$ satisfy the commutation relations of the original $Z$ and $X$ spins respectively, and $\zeta_{B/A}$ commutes with $\xi_{A/B}$. In this dual language, the edge Hamiltonian for n-periods reads:
\begin{align}
H_\text{eff}^{(n)} = \sum_{j,k=1}^{n-1}\sum_i h_{2i}^{(k)}\(\zeta_{A,i}^\dagger \zeta_{A,i+1}\)^{jk}+\sum_{k=1}^{n-1}h_{2i-1}^{(k)}\(\xi_{A,i}\)^{k}+\[A\leftrightarrow B, (2i)\leftrightarrow (2i-1)\]+h.c.
\end{align}
\end{widetext}

Again, we find that the, in the dual language, the $A$ and $B$ $\Z_n$-chain couplings are related by exchanging spin-exchange terms of $A$ for transverse field terms of $B$. Then if the $A$ sublattice lies at the critical point between the $\Z_n$ ordered and disordered phases (in the dual language), so too will the $B$ sublattice spins. In the original spin language, this corresponds to the edge being in a symmetry preserving, quantum critical state, which can be interpreted as the critical point between the 1D SPT and trivial phases. Note however, unlike the $n=2$ case, there are $n$ spin-exchange terms, and $1$ transverse field terms, so the statistical translation invariance of $h_i$ is no longer sufficient to ensure the criticality of the edge. Rather, the even and odd fields must be fine tuned to achieve the critical edge termination for $n>2$.

More generally, we expect the edge of this model to exhibit localized and spontaneously symmetry breaking phases for strongly disordered edge couplings, $h_i^{(k)}$. Or to exhibit a thermal phase for weakly disorder edge couplings. Nonetheless, even in the infinite temperature thermal edge, a sign of the underlying SPT order persists. Namely, if one considers the system on a cylinder, and inserts a $\Z_n^A$ symmetry flux through the cylinder axis. Then the $\Z_n^B$ charge of each edge state is incremented by one unit for each Floquet period, returning to its original value only after $n$ cycles. This feature is also present even when the edges are thermal and at infinite effective temperature. 

\bibliography{FloqSPTbib}

\end{document}